\documentclass[article]{aa}

\usepackage{graphics}
\usepackage{epsfig}
\usepackage{latexsym}
\begin{document}
\hyphenation{brems-strah-lung}
\title{Type-I  bursts within outbursts of IGR~J17473-2721}

\author{Yu-Peng Chen\inst{1}, Shu Zhang\inst{1}, Diego F. Torres \inst{2}, Jian-Min Wang\inst{1,3},        Ti-Pei Li\inst{1,4}}

\institute{Laboratory for Particle Astrophysics, Institute of High
Energy Physics, Beijing 100049, China
\and
ICREA \& Institut de Ci\`encies de l'Espai (IEEC-CSIC), Campus UAB,
Facultat de Ci\`encies, Torre C5-parell, 2a planta, 08193 Barcelona,
Spain
\and
Theoretical Physics Center for Science Facilities (TPCSF), CAS
\and
Center for Astrophysics,Tsinghua University, Beijing 100084, China
          }

\offprints{Yupeng Chen}
\mail{chenyp@mail.ihep.ac.cn}

\date{ }

\date{Received  / Accepted }

\titlerunning{Type-I  bursts within outbursts of IGR~J17473-2721}
\authorrunning{Yupeng Chen et al.}

  \abstract
  {}
{Two outbursts were observed by RXTE in the history of the atoll source IGR~J17473-2721. During the most recent outburst in 2008, the source showed a complete series of spectral states/transitions. The neutron star system was prolific in type-I X-ray bursts, and we investigate them in the context of complete outbursts evolution. }
{A total exposure of $\sim$ 309 ks was collected by RXTE during the two outbursts of IGR~J17473-2721. We carried out a systematic search for type-I bursts in this data set.
 For each burst found, we investigated the burst profile, the peak flux, and their dependence on the accretion rate along the evolution of the outbursts. }
{Eighteen type-I X-ray bursts were found from IGR~J17473-2721: two from the outburst in 2005 and the other 16 from the recent outburst in 2008. Among them,  3 bursts show photospheric radius expansion (PRE). The distance to the source is estimated as 6.4 kpc with a 15$\%$ uncertainty  based on the three bursts that show PRE. In the recent outburst, there are 6 bursts showing up in the low/hard state prior to the state transition to a high/soft state, 3 bursts at the end phase of the high/soft state, and 7 in the following low/hard state. The blackbody radius of these bursts  presents a variety of interesting features.
We find that, at the end of the recent outburst, the profile of the  blackbody radius is anti-correlated with the blackbody temperature and the burst flux. The durations of the type-I burst are found to correlate with the Eddington ratio and to have two parallel evolution groups. Along the decreasing Eddington ratio, the burst duration decreases and ends in each group the PRE bursts occurred. This provides new clues to the type-I bursts in the context of outbursts for atoll XRBs.}
{} 

   \maketitle

\section{Introduction}
Type-I X-ray bursts were discovered during the mid-1970s
(Grindlay et al. 1976; Belian et al. 1976),
and manifest as a sudden increase (typically a factor of 10 or greater) in the X-ray
luminosity of  neutron star systems (NSs).
Typical type-I X-ray bursts exhibit fast rise times ($\la 1$ and $10$~s),
and slower exponential decay (tens to hundreds of seconds).
 These bursts are caused by unstable  burning of
accreted H/He on the surface of neutron stars in low-mass X-ray
binary (LMXB) systems, in contrast to type-II bursts,
which are thought to be caused by an increase in the accretion rate from the companion `donor' star.
For reviews, see
Lewin et al. (1993), Cumming (2004), and Strohmayer \& Bildsten (2006).
Recently,
Galloway et al. (2008)
has
presented a large
sample of bursts (1187 in total) observed by {\it RXTE} over a time interval
of more than ten years.

Atoll sources have a lower accretion rate than Z sources  and can have hard states, soft states, and  transitions between them. In the hard state, a significant fraction
of the energy is emitted above 20 keV and is normally accompanied by
radio jets.
The source in the soft
state has relatively much weaker emission above 20 keV and is believed
to be dominated by  the accretion disk and the boundary
layer where the accreted matter impacts the NS surface.
For reviews, see
Hasinger \& van der Klis (1989) and van der Klis (2006).
Bursts are more prolific in atolls than in Z sources and have been observed in all of the
above X-ray states (e.g., Galloway et al. 2008).

IGR~J17473-2721 has been identified as an atoll source (Altamirano et al. 2008b).  To date, IGR~J17473-2721 has shown
two outbursts observed by RXTE,  in 2005 and a more recent one in 2008. The 2005 outburst was found to remain  in the low/hard state, while the recent one consists of a variety of spectral states/transitions. The source stayed in the low/hard state for 2 months and then transferred to  the high/soft state in 3 days; during the decay, the source transferred back  to a low/hard state at a lower luminosity level, forming the so-called hysteresis.  The spectral analysis may suggest  a  highly   inclined XRB system (Zhang et al. 2009). The long-lived preceding low/hard state makes IGR J17473-2721 resemble the behavior of outbursts seen in black hole X-ray binaries like GX 339-4. For details of RXTE observations on the two outbursts see Zhang et al. (2009). Several type-I bursts were detected in IGR~J17473-2721  during the outbursts:
2 type-I X-ray bursts in 2005 were reported  by Galloway et al. (2008).
Several other bursts detected in the 2008 outburst were first reported by Del Monte et al. (2008), with SuperAGILE, and then by Altamirano et al. (2008a) with {\it SWIFT}. Altamirano et al. (2008b) estimate the distance is likely to be between 4.9 kpc (using the theoretical Eddington limit for a pure He atmosphere) and 5.7 kpc (using the empirical value for the Eddington luminosity from Kuulkers et al. 2003) based on the peak flux of two bursts that show photospheric radius expansion.

Given that IGR~J17473-2721 is among the few atoll sources that show a large number of type-I bursts occurring in the context of complete outbursts  (Zhang et al. 2009), we carried out the analysis on each type-I burst, based on the most updated RXTE observations and analysis tools.
 The observation and data analysis are commented on in Section 2, the results in Section 3, and finally the discussion and summary in Section 4.

\section{Observations and data analysis}
Public data from {\it RXTE~} of IGR~J17473-2721 include 130 {\it RXTE}/PCA pointed observations, with the identifier (OBSID) of proposal number (PN) 91080 and 93442, in the data  archive of the High Energy Astrophysics Science Archive Research Center (HEASARC).
These observations sum up $\sim$ 309 ks of exposure on the source and scatter over the short outburst in 2005 and the long outburst in 2008
(Zhang et al. 2009).
The analysis of PCA data was performed by using
Heasoft v. 6.2.  We filtered the data using the standard \emph{RXTE}/PCA criteria.
To be precise, only the PCU2 (in the 0-4 numbering scheme) was used for the analysis, because only the PCU2 was 100\%  during the observation,
and we picked up the time intervals under the following constraints on elevation angle $>10^{\circ}$, and pointing offset $<0.02^{\circ}$. Elevation is the angle above the limb of the Earth, and the pointing offset is the angular distance between the pointing and the source. The background file used in the analysis of PCA data is the most recent one available on the HEASARC website for bright sources\footnote{
pca$\_$bkgd$\_$cmbrightvle$\_$eMv20051128.mdl},
and the detector breakdowns have been removed.\footnote {See the website

http:$//$heasarc.gsfc.nasa.gov$/$docs$/$xte$/$recipes$/$pca$\_$breakdown.html
for more information.}
An additional  1\% systematic error was added to the
spectra because of calibration uncertainties, if not otherwise specified.  The
spectra were fitted with XSPEC v12.3.1 and the model parameters were
estimated  with 90$\%$ confidence level.

\section{Results}

\begin{figure}[ptbptbptb]
\centering
 \includegraphics[angle=0, scale=0.45]{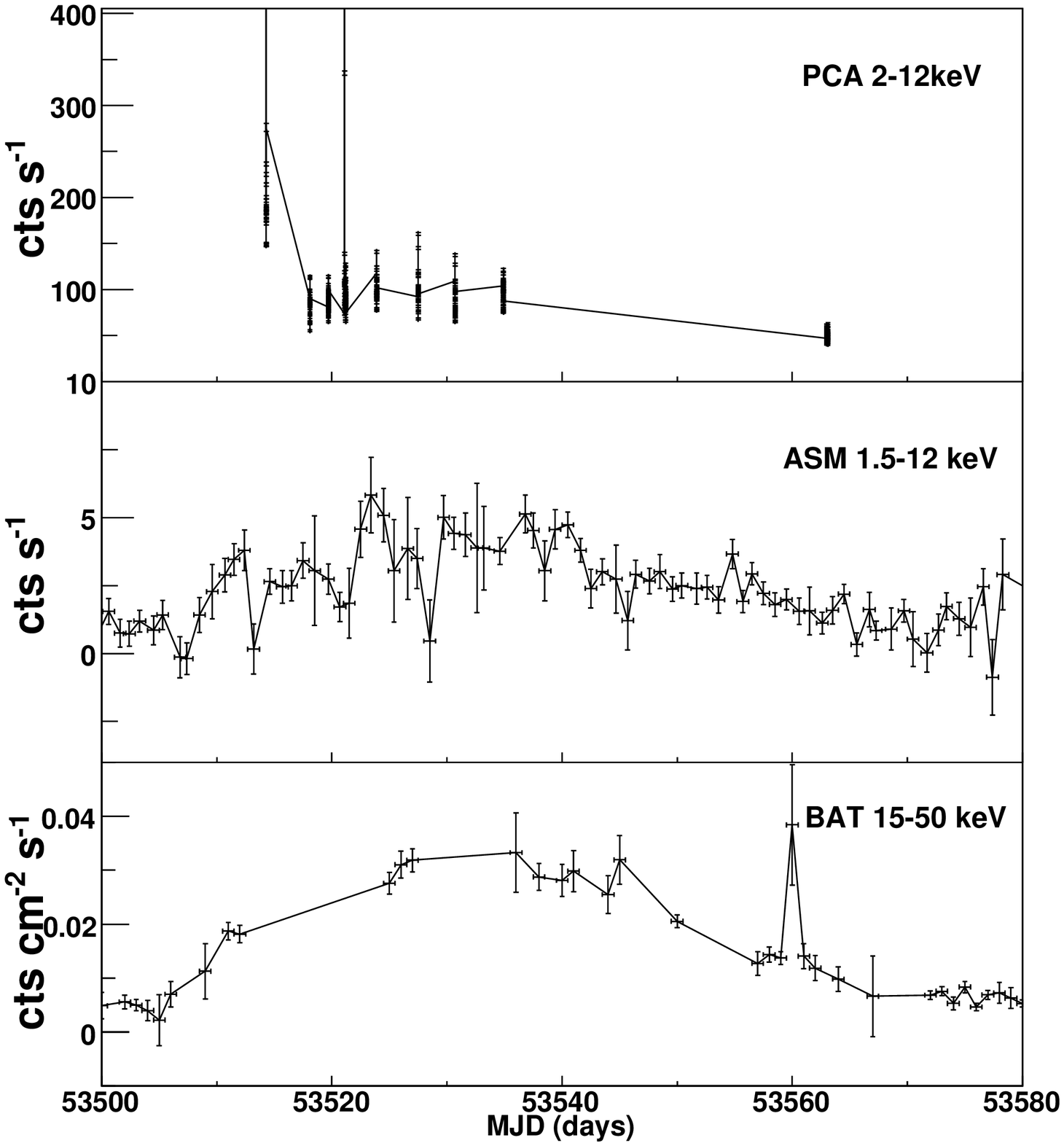}
  \includegraphics[angle=0, scale=0.452]{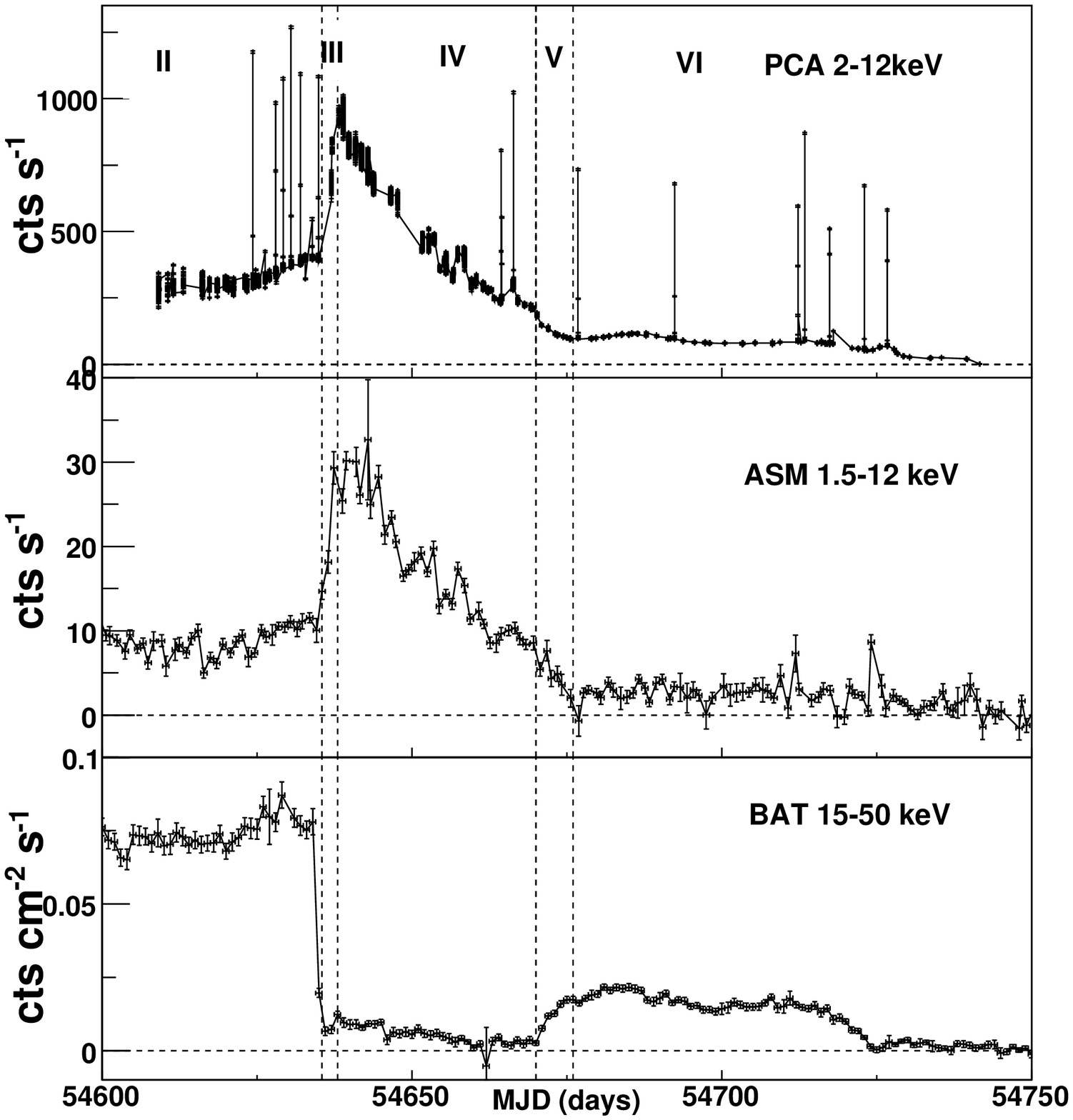}
      \caption{
       The PCA(PCU2)/ASM/BAT lightcurves of IGR J17473-2721 during the short outburst (left panel) in 2005 and the recent outburst (right panel) in 2008. The type-I X-ray bursts are the spikes in the PCA lightcurve.}
         \label{lc_flux}
\end{figure}

We carried out a systematic search for bursts in the lightcurve of  IGR~J17473-2721. The lightcurve was extracted  in 64 seconds time bins for every observation identifier (OBSID) with PCA standard 2 model for each energy channel.
The lightcurves of the two outbursts are  shown in Fig. \ref{lc_flux}, where the background has been removed. For comparison, we also put together the ASM/RXTE (1.5-12 keV) and BAT/SWIFT (15-50 keV) lightcurves for the two outbursts. The entire 2008 outburst was divided into 6 regions according to the spectral state (following Zhang et al. 2009): regions II/VI for low/hard states, region IV for high/soft state, and regions III/V for the state transitions.  The type-I X-ray bursts appear in the PCA lightcurves as spikes, with 2 detected in the 2005 outburst and 16 in the 2008 one.  A lightcurve
with a time binning of 16 seconds shows no additional shorter bursts.

\begin{figure}[ptbptbptb]
\centering
 \includegraphics[angle=0, scale=0.5]{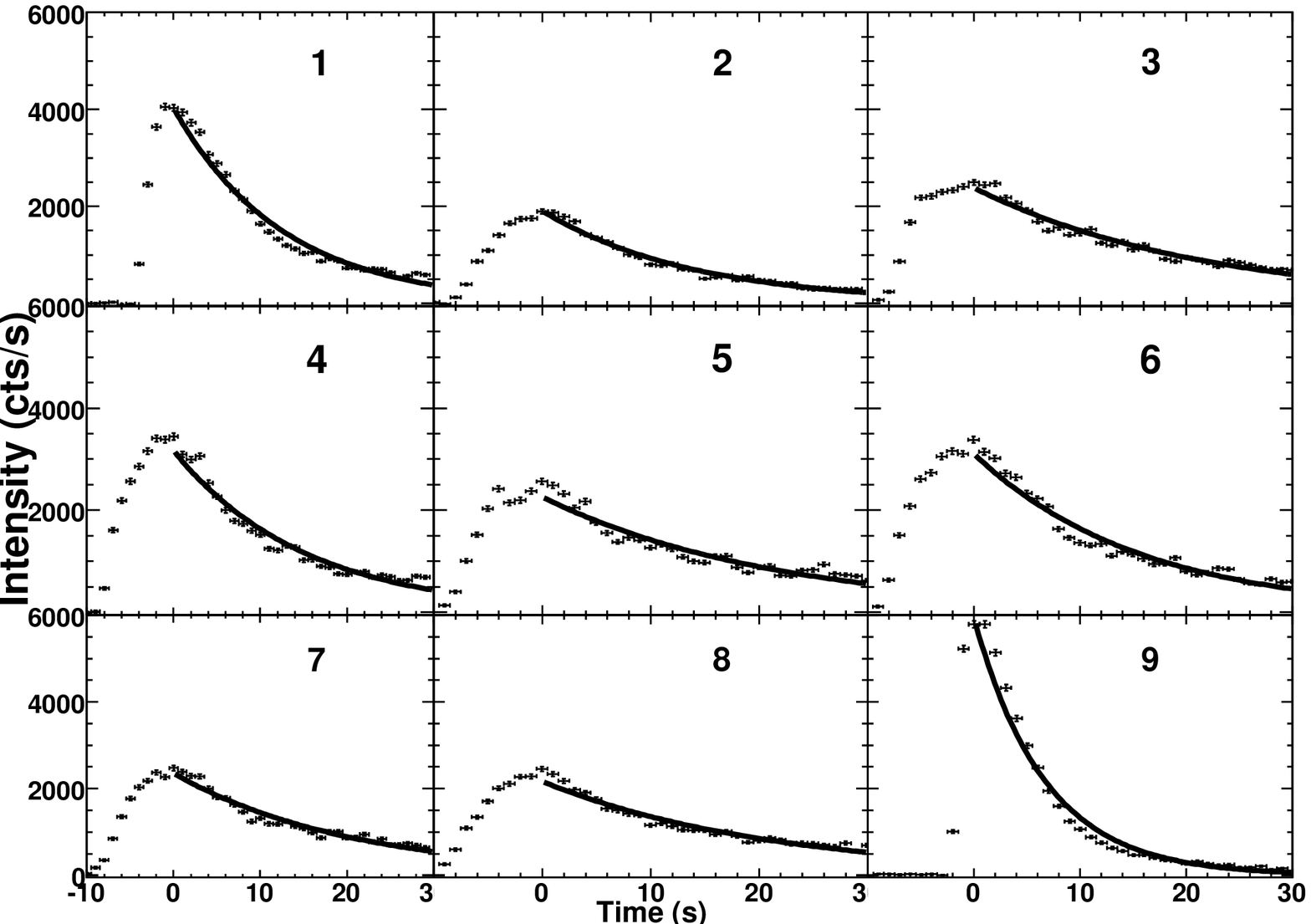}
  \includegraphics[angle=0, scale=0.5]{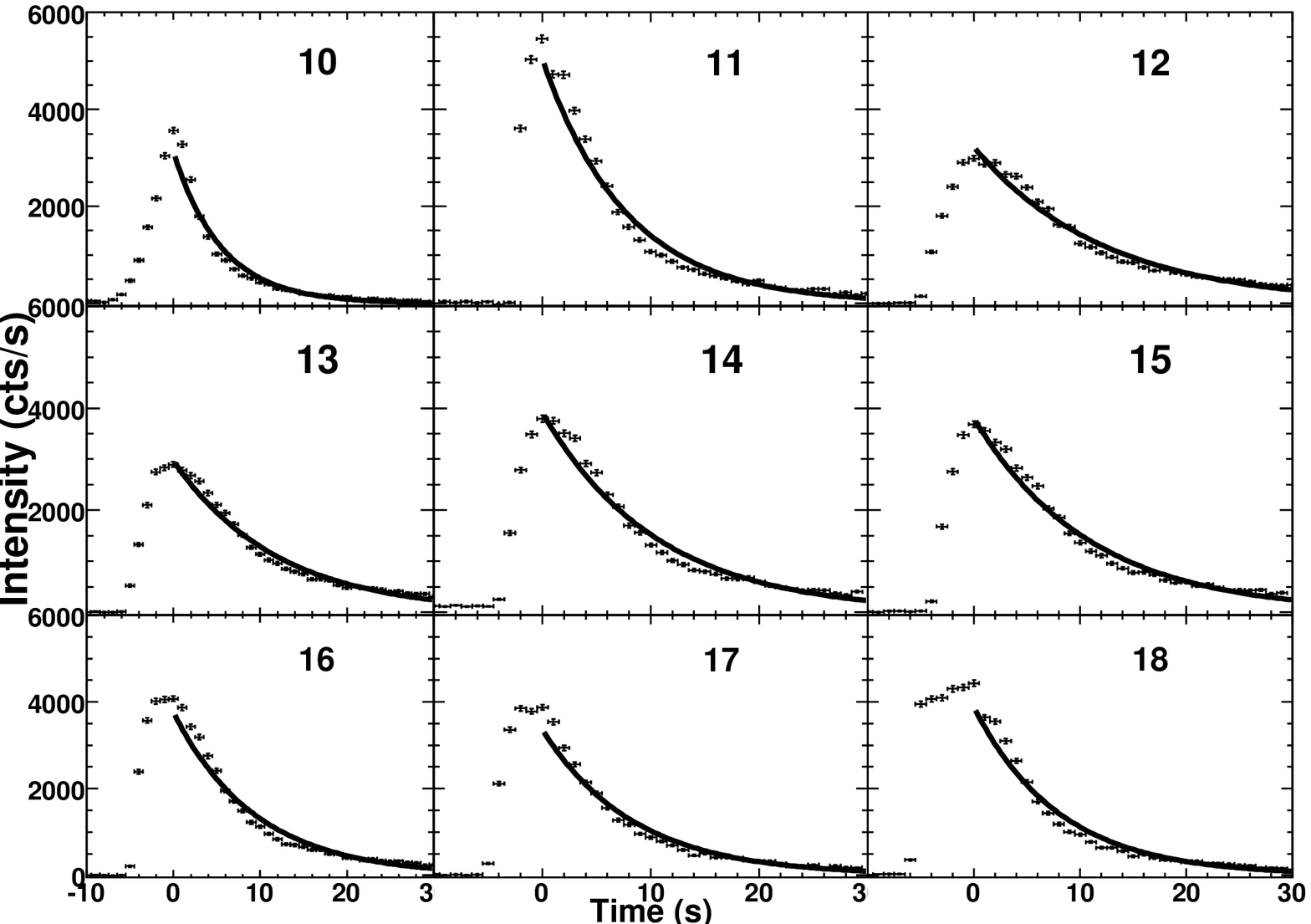}
      \caption{The profiles and the fit to their decays with an exponential shape for the 18 bursts. The time bin is 1 second.}
         \label{lc_burst}
\end{figure}

The profile of each  type-I X-ray burst was derived in time bins of 1 second under PCA E
mode  (E$\_$125us$\_$64M$\_$0$\_$1s, which has 125 $\mu$s time resolution in 64 energy channels starting at channel 0, 1 second read out time) in  the 2-60 keV band (Fig. \ref{lc_burst}).  The background and persistent emission were taken off.
For each burst, we estimated the peak time  and the peak flux from the lightcurves, and  the decay time from fitting the decay part of the lightcurve with an exponential shape (Fig. \ref{lc_burst}). The rise time is defined as the interval  during which the flux  increases  from $10\%$ to $90\%$ of the peak value (Galloway et al. 2008).
These parameters are presented in Table 1, for each of the bursts. The hardness ratio, defined as the flux ratio   (6-30 keV)/(2-6 keV), is computed for each burst in 0.25 second time bins (Fig. 3), which is used to investigate the burst evolution (Strohmayer \& Bildsten, 2006). Accordingly, 3  bursts (\#  9, 11, 18) are found to be the PRE events.

The spectra of each burst were extracted with PCA E mode mostly in 3-20 keV for every 0.25 seconds, based on only PCU 2 data. We took the BBodyrad model to fit the data, with the absorption fixed at 3.8$\times$10$^{22}$ atoms/cm$^2$ (Altamirano et al. 2008a).
 We used the spectral data  at 100 seconds  prior to the burst as background,
subtracting it from the burst data  (as in Muno et al. 2000).
The data of the persistent emission
are fitted with a model of disk blackbody plus power law, commonly used to model atoll sources,
under a fixed column density 3.8$\times$10$^{22}$ atoms/cm$^2$ as derived in  Altamirano et al. (2008a)
and a fixed  iron line at 6.4 keV.
The reduced $\chi^2$ of the fit for each burst are generally
$\sim$0.8--1.2, mostly showing  good fits. The resulting evolution of the blackbody temperature (T$_{bb}$) and the corresponding radius (R$_{bb}$) of the emission area are plotted in Fig. 4.  The values of T$_{bb}$, R$_{bb}$, and the luminosity F$_{pk}$ derived at the peak flux, the persistent luminosity F$_{per}$ prior to each burst, and the bolometric luminosity E$_b$ calculated in the energy band 1.5-30 keV by summing up the luminosity over the entire burst are shown in Table 1. Enclosed as well is the characteristic timescale $\tau$ for each burst, derived as E$_b$/F$_{pk}$.

As shown in Fig. 4, the shape of the evolution of  T$_{bb}$ in time always follows the evolution in the burst flux except for the three PRE bursts (\# 9, 11, 18), where the R$_{bb}$ curve shows a variety of trends. The  R$_{bb}$ curve   monotonically increases along the evolution  of the bursts \#  1 and 10, showing that the burning expands across the surface of the neutron star. The bursts \#  9, 11, and 18 have an R$_{bb}$ peak, which is typical of PRE bursts. It is interesting  that there are 4 bursts (\# 14, 15, 16, and 17) at the end of the outburst in 2008 that show an anti-correlation between R$_{bb}$ and the burst flux. The peak temperatures of these bursts are about 3 keV, higher than that averaged over the others (about 2.5 keV). It seems that the burning area in these bursts shrinks with  the increasing flux and spreads with the decreasing flux.  Actually, such a feature was reported in one burst observed from  KS 1731-260 (Muno et al. 2000) and was studied further  by Bhattacharyya et al. (2009), using the sample of type-I bursts derived in  Galloway et al. (2008). They find an anti-correlation between the burst decaying flux and the blackbody radius for short type-I bursts. The remaining 9 bursts have roughly constant R$_{bb}$, indicating less change of the burning area across the burst.

To further investigate the burst behavior, we followed the analysis done in
Galloway et al. (2008) on the RXTE type-I X-ray bursts.
The diagrams of the persistent flux (at 1.5-30 keV) represented  by the Eddington ratio $\gamma$=L$_{bol}$/L$_{Edd}$  (L$_{Edd}$ takes $3.79\times 10^{38}$ ergs s$^{-1}$ from Kuulkers et al. 2003)
 are shown in Fig. 5,
and can be indicative of the accretion rate vs the time scale $\tau$. We find that $\tau$ is correlated with the  Eddington ratio. Along the outburst, the $\gamma$-$\tau$ evolutions appear in two groups: along the decreasing Eddington ratio, the burst duration decreases and ends up in each group when the PRE bursts occur. A linear fit to each group shows that the slopes for the two groups are consistent within errors:  $419\pm19$ for the group at higher $\gamma$ and $469\pm131$ for the group with $\gamma$  less than 0.05.

   The accretion rate can be better estimated from the position in the color-color diagram (CCD) (van der Klis 1995)
    by introducing a parameter so-called $S_a$, calculated via the projection to the evolution  appears in CCD (M\'{e}ndez et al. 1999).
We calculated PCU2  count rates in four energy bands (3-4 keV, 4-6.4 keV, 6.4-9.7 keV and 9.7-16 keV), with the background  subtracted. We define the flux ratio  (4--6.4 keV)/(3--4 keV) as the soft color
 and the flux ratio (9.7--16 keV)/(6.4--9.7 keV) as the hard color. In Fig. 6
we show a CCD for the two outbursts of IGR J17473-2721, with each point representing 64 seconds of data.
The color-color positions of each of the bursts are plotted in Fig. 6.
We use the parameter $S_{a}$ to measure the position along the evolution of persistent flux.
We set  $S_{a}$ to 1 at CCD coordinates (2.00,0.78), and to 2 at (1.41,0.40).
For simplicity, we approximate the shape of the color-color diagram with a three-segment broken line.  The data  close to each burst are projected upon the track,
 and the derived $S_a$ are fitted a Gaussian distribution to estimate the mean   $S_{a}$ of each burst. The  $S_{a}$-$\tau$ diagram is then shown in Fig. 7. We find that the $\tau$ is anti-correlated with the $S_{a}$ for  the group with large $\gamma$ in the $\gamma$-$\tau$ diagram (see Fig. 5), indicating an increasing accretion rate. For the group with small $\gamma$, most of $S_{a}$ are around 1.2 and show no obvious evolution in accretion rate except for the last two bursts \#  17 and 18. The CCD positions around  these two bursts are probably returning to  the banana region (see Fig. 6).
We investigated  these two bursts further via timing analysis, and derived the evolutions of the soft and hard color, and the lightcurve in different energy bands (see Fig. 8 and 9). We  find that, while the   soft color evolves consistently,  there is a dip for the hard color within  MJD 54720-54730, covering bursts \# 17 and 18 (Fig. 8). The lightcurves show that during the decay of the flux at hard X-rays (above $\sim$ 10 keV), an
additional flare appears at soft X-rays (below $\sim$ 10 keV), which leads to a dip in hard color (Fig. 9).
We speculate that this additional soft flare  might correspond to a flickering of the accretion rate in the disk
and that happened on top of the overall decay of the outburst.

\section{Discussion and summary}

During the two outbursts showing up in the observed   history of IGR J17473-2721, 18 type-I bursts have been discovered by RXTE in the 3-20 keV band. Apart from the 16 type-I bursts reported by
Altamirano et al. (2008b), 7 bursts (\# 12-18) are found here at the end phase of the recent outburst. These bursts are investigated in detail within the context of the entire evolution of the outburst, and several features are found.

The rise and decay times  of the first 6 bursts (\# 3-8)  are in general  longer than those of the last 7 bursts (\# 12-18) in the recent outburst. In both cases the source was in a low/hard state. The shortest time scales of the rise and the decay come from the three bursts (\# 9, 10, 11) occurred at the end of the high/soft state of the recent outburst.  The Eddington ratios in these three time regions are about 0.1L$_{Edd}$ (region II, bursts \# 3-8), 0.07L$_{Edd}$ (region IV, burst \# 9-11) and $<$ 0.05L$_{Edd}$ (region VI, bursts \# 12-18). It is obvious that the corresponding burst rate in the low/hard state (region II) prior to the transition to high/soft state is higher than in the remaining part of the outburst (regions III-VI).
 The net exposure is $\sim$ 29 ks covering the 6 bursts (\# 3-8) in the time interval II (the low/hard state prior to the transition to the high/soft state), and $\sim$ 72 ks enclosing  the 7 bursts (\# 12-18) in the time interval VI (the low/hard state transferred from the high/soft state).
This is consistent with the theoretical prediction for a higher occurrence rate of bursts with increasing accretion rate.

 It is thought that  burst properties depend on the mass accretion rate and the composition of the accreted matter ($\dot{m}$)
(e.g. Galloway et al. 2008). Since it is given by a He accretor, the burst will always be short irrespective of the accretion rate, which is not the case in IGR J17473-2721.  An exception to He accretor is that, for low accretion rates and therefore cool envelopes, He can ignite in a thick layer, and power bursts on timescales of as long as one hour (in 't Zand et al. 2009).
With a mixed H/He accretor, as might be the case for IGR J17473-2721, the theory predict a correlated accretion rate and burst duration. With a higher accretion rate, the temperature in the fuel layer becomes hotter, leading to an earlier H ignition (Fujimoto et al. 1981).
 As long as the accretion rate is high enough for
hot CNO cycle burning of the accreted H, since the CNO cycle is $\beta$ limited,
the conversion rate from H to He remains the same, so that less H is consumed before each
burst. As the accretion rate goes up, the fraction of H in the fuel
also goes up. This can then account for the increased duration, since it is the
  H burning via the rp-process that leads to longer durations.
At low enough accretion rates, there will be
a rate at which most the accreted H is processed before ignition. This
will lead to He-dominated, shorter bursts.
Similar correlation between accretion rate and burst duration has been reported in EXO 0748-676 (Galloway et al. 2008).  Here we see that trend for IGR J17473-2721 in the luminosity range 0.01-0.1L$_{Edd}$, if the  Eddington ratio can be the proper indicator of accretion rate: the rise and decay of the bursts are longer at higher values of the Eddington ratios. This is indicated as well in Fig. 5, the plot of the Eddington ratio $\gamma$ vs the characteristic duration $\tau$, which shows a good correlation.

Along decreasing Eddington ratios, the burst duration becomes shorter, and ends up with the occurrence of the PRE bursts (\# 9, 11, and 18), which have the shortest duration.  However, we notice that in Fig. 5 such a $\tau$-$\gamma$ correlation  happens twice.  After the end of the high $\gamma$ group, the burst jumps to a  higher $\tau$ in the low $\gamma$ group, which is hard to understand by considering that the longer burst duration is  caused solely by a  higher accretion rate. A possible explanation might be that the bursts are mixed H/He burns, and the H/He fraction can be different in the PRE burst and the subsequent one even if they have the similar Eddington ratio.
 An alternative choice might be that the PRE bursts might not consume all the He fuel. As shown in Weinberg et al. (2006), the PRE burst is launched at a thin
layer at the base of the accreted fuel, since the nuclear energy generation is sensitive
to both temperature and density. Above this thin layer is a convective region, where
the fuels are consumed. After the PRE burst, both the ashes in the convective
region and the fuels outside of this convective region might return to the neutron star surface,
contributing to the fuel for the next burst. Also some ashes of the burning are ejected by a radiation-driven wind, and such mass ejection could expose the He-rich material rather than H
(Weinberg et al. 2006).
It remains an open question why the PRE bursts occur at the lowest accretion rate at the end of both evolution groups shown in Fig. 5.

 The parallel character of the two burst groups  in Fig. 5 recalls what shows up in the QPO-luminosity diagram (M\'{e}ndez et al. 1999), which suggests that the X-ray flux  is  not always a good indicator of accretion rate. The bursts (\# 9, 10, 11, and 18) at the end of both groups are located in the banana region of the CCD, with higher $S_a$ values compared to the others that  show no clear trend to evolution of accretion rate. For the burst group with higher $\gamma$, the accretion rates   within the bursts \#  9, 10, 11 are  larger   than in the earlier ones, but the burst duration goes down. This is, however, inconsistent with the scenario that the accretion rate and the burst duration can be correlated (Fujimoto et al. 1981).  A similar case was reported by Muno et al. (2000) on one burst event of KS 1731-260, where the fast burst occurs at a relatively high accretion rate. As pointed out by Muno et al. (2000), such a paradox can be solved by taking the accretion rate per unit area into account at the surface of the neutron star, which is reduced for bursts with a larger burning area (Bildsten 2000).

Actually, we find that, for the burst group with higher $\gamma$ values, the apparent blackbody radius of the bursts \#  9, 10, and 11 are larger than those of the earlier ones.  Such a trend is also visible in burst \# 18 for the burst group with low $\gamma$.
 The large R$_{bb}$ means  low real local accretion rates, hence low luminosities if the soft X-rays are not dominated by the emission from the accretion disk.
Actually, our spectral analysis of outburst from IGR J17473-2721 (Zhang et al. 2009)
does show that the soft X-rays are not dominated by the emission from an accretion disk, probably due to
the high inclination of the disk.  The two groups of the bursts, as shown in Fig. 5 and Fig. 7, happened in the low/hard state prior to the transition to the high/soft state  and in the low/hard state following the high/soft state, respectively. Therefore, they correspond to two different kinds of the low/hard states
(the so-called hysteresis)  along the outburst of IGR J17473-2721.
We notice that the persistent emission around bursts \# 17 and 18 have special color evolutions.
While the soft color evolves almost continuously, the hard color shows a narrow dip ($<$ 10 days)
 around these two bursts.
Such dip is caused by an additional soft flare happen sporadically on top of the decay of the outburst, which may have led to the deviation in CCD or $S_a$-$\tau$ diagram from the overall tracking of the accretion rate for the entire outburst.

PRE bursts are those where
the peak flux reach  levels comparable to
the Eddington luminosity at the surface of the NS. The
 radiation pressure exceeds the gravitational force
binding the outer layers of accreted material to the star.
Such bursts frequently exhibit a characteristic spectral evolution in the
first few seconds, with a local peak in blackbody radius and
a dip in temperature at the same time, while the flux remains
approximately constant. This pattern is thought
to result from expansion of the X-ray emitting photosphere once the burst
flux reaches the Eddington luminosity. The effective temperature must
decrease in order to maintain the luminosity at the Eddington limit, and
excess burst flux is converted into kinetic and gravitational potential
energy in the expanded atmosphere.
Consequently, the PRE bursts can be used
as distance  indicators (Basinska et al. 1984).  So far, three such PRE bursts have been detected  in IGR J17473-2721, with  a peak flux measured within a range of 6.92-8.23 $10^{-8}$ erg cm$^{2}$ s$^{-1}$.
We therefore derive a distance of 6.4 kpc with the average of the three peak flux 7.78 $10^{-8}$ erg cm$^{2}$ s$^{-1}$ using the empirically determined Eddington luminosity $3.79\pm0.15 \times 10^{38}$ erg\,s$^{-1}$ for PRE
burst (Kuulkers et al. (2003) with an uncertainty of $15\%$).  This is consistent with the previous estimation of the distance with an upper limit  of 6.4 kpc by
Galloway et al. (2008).

Apart from the three PRE bursts that show  the typical evolution of T$_{bb}$ and R$_{bb}$ in Fig. 4, the 15 other bursts have the T$_{bb}$ profiles that can be naturally interpreted as heating resulting from the initial fuel ignition, followed by cooling of the ashes once the available fuel is exhausted. However, the 15 bursts have a variety of evolution trends in R$_{bb}$, which may indicate that the non-PRE bursts can require more complicated procedure than the PRE ones. Along the burst, the burn can be restricted to an area on the surface of the neutron star, spread steadily along the burst flux evolution, or have an anti-correlation between the burning area and the burst flux. Recently, a very rare triple-peaked burst event was discovered in XRB 4U~1636-53 (Zhang G. B. et al. 2009). These authors find that, along the burst evolution, the radius of the burning area increases steadily and shows two dips when the flux got its second and third peaks. These results show that the various trends in radius evolution, which are similar to those found in IGR J17473-2721, can even co-exist within one type-I burst.

 We notice that
the anti-correlation between the blackbody radius and the decaying
flux was seen previously in one burst of KS 1731-260 (Muno et al., 2000).
Bhattacharyya et al.
(2009) carried out the most recent investigation of such a relation,  using a sample of type-I bursts derived in
Galloway et al. (2008). They find an anti-correlation between the
burst decaying flux and the blackbody radius for short type-I burst, but a positive
correlation for a long-duration type-I burst. A possible explanation for this discovery
is related to a spectral hardening factor (or the so-called color factor),
which accounts for hardening due to the scattering of the photons by the electrons
in a neutron star atmosphere. Such a color factor is a function of the chemical composition
of the neutron star atmosphere, the actual surface temperature, and the stellar
surface gravity. The first two elements could be related to the type-I burst of the
different types, He or H/He dominance. Therefore, a change in these may affect the apparent
blackbody radius measured in the burst. For IGR J17473-2721, we show at least 4 samples of such a
type-I event with anti-correlation between decaying flux and blackbody radius. Their
burst durations are shorter than most of the others, which show no  trend in correlation
between decaying flux and blackbody radius. This is consistent with the discovery
in Bhattacharyya et al. (2009) from a sample of 900 bursts from 43 sources.
Another possibility is that, since the persistent emission is known to vary, it is possible
that some trends seen in inferred radius could be associated with incomplete modeling/subtraction of the persistent emission.  A similar issue has been discussed by Muno et al. (2000) in their
handling of the background subtraction of the type-I burst of KS 1731-260. There, these authors also adopted the persistent emission 100 seconds prior to the burst as the burst background, and ignored the possible variability of the persistent  emission during the burst. The possible variability of the persistent  emission in each burst, if any, is hard to disentangle  from the overall burst emission.

In summary, the 18 type-I X-ray bursts discovered during two complete outbursts of IGR J17473-2721 provide us some new insights into the burst evolution. In particular, the discovery of two similar evolution groups in duration/Eddington ratio diagrams,  both ending with the occurrence of the PRE burst at the lowest Eddington ratio, challenges the current modeling of the type-I bursts of XRBs.

\acknowledgements
We are grateful to the anonymous referee for his/her comments, which were
helpful in polishing the paper.
This work was subsidized by the National Natural Science Foundation of China,
the CAS key Project KJCX2-YW-T03, and 973 program 2009CB824800. J.-M. W. thanks
the Natural Science Foundation of China for support via NSFC-10325313, 10521001,
and 10733010. DFT acknowledges support from grants AYA2009-07391 and SGR2009-811.
This research made use of data obtained
through the High Energy Astrophysics Science Archive Research Center Online Service, provided by the NASA/Goddard
Space Flight Center.

\onecolumn

\begin{table}[ptbptbptb]
\begin{center}
\label{tab_burst}
\caption{The 18 type-I X-ray bursts' rise time, decay time, the temperature of BB, and the radius, bolometric flux, fluency, characteristic time scale $\tau$, and bolometric flux of the persistent flux.
 }
\begin{tabular}{cccccccccccccccccc}

\hline \hline
 No& OBSID & Time & Intensity                &$t_{r}$   &$t_{d}$   &$T_{bb}$  &$R_{bb}$   &$F_{pk}$   &$E_{b}$   &$\tau$ &$F_{per}$\\
    &       & (MJD) &$(cts~s^{-1}$)                  &(s)        &(s)      &(keV)     &(km)       &($10^{-8}$) &($10^{-7}$)   & (s)&($10^{-9}$)\\\hline
1& 91050-02-01-00 &53514.297601 &4055&4&12.7&2.46$^{+0.11}_{-0.11}$  &$7.7^{+0.7}_{-0.6}$ &6.0 &12 &20.0  &3.30\\\hline
2 &91050-02-05-00 &53521.108180 &1896 &6&13.9&$2.25^{+0.15}_{-0.14}$  &$6.0^{+0.8}_{-0.7}$&2.56 &5.36&20.9  &2.57\\\hline
3 &93442-01-03-01 &54624.293851  & 2498&6&21.7&$2.33^{+0.14}_{-0.13}$  &$6.6^{+0.8}_{-0.7}$&3.61 &10  &27.3&7.42\\\hline
4 &93442-01-03-05 &54628.026895  &3443 &6&15.1&$2.76^{+0.17}_{-0.16}$  &$5.7^{+0.6}_{-0.5}$&5.01 &10.64& 27.7&8.00\\\hline
5 &93442-01-03-06 &54629.213273  &2562 &5&21.5&$2.54^{+0.17}_{-0.16}$  &$5.7^{+0.8}_{-0.6}$&3.74 &9.12&21.2  &8.02\\\hline
6 &93442-01-04-00 &54630.523574  &3378 &5&15.6&$2.79^{+0.17}_{-0.16}$  &$5.5^{+0.7}_{-0.6}$&5.15 &10.72& 20.8 &8.85\\\hline
7& 93442-01-04-01 &54632.008828  &2473 &6&20.6&$2.35^{+0.16}_{-0.15}$  &$6.4^{+1.0}_{-0.7}$&3.54 &9.12 &25.8 &9.25\\\hline
8 & 93442-01-04-04&54634.909141  &2457 &7&21.4&$2.53^{+0.20}_{-0.18}$  &$5.6^{+0.9}_{-0.7}$&3.51 &9.12&26.0 &9.21\\\hline
$9^{*}$&  93442-01-08-06&54664.351270  & 5789 &2&6.7 &$2.32^{+0.10}_{-0.09}$&$10.1^{+0.8}_{-0.7}$&8.23 &7.36 & 8.9 &5.42\\\hline
10&93442-01-08-06& 54664.486930 & 3563 &6&5.6 &$2.41^{+0.13}_{-0.12}$&$7.3^{+0.8}_{-0.7}$&4.99 &4.08 & 8.18 &5.15\\\hline
$11^{*}$& 93442-01-09-00& 54666.359905 & 5058 &2&7.8&$2.35^{+0.09}_{-0.09}$  &$9.8^{+0.8}_{-0.7}$&8.18 &6.72 & 8.21&5.86\\\hline
12& 93442-01-10-05& 54676.784407&2989&4&12.2&$2.36^{+0.12}_{-0.11}$  &$7.0^{+0.7}_{-0.6}$&4.16 &7.2 &17.3&2.61\\\hline
13& 93442-01-12-05& 54692.425935  &2898 &4&12.0&$2.41^{+0.14}_{-0.13}$  &$6.6^{+0.8}_{-0.6}$&4.11 &6.8  &16.5&2.50\\\hline
14& 93442-01-15-04& 54712.285321  &3792 &3&10.6&$2.76^{+0.16}_{-0.15}$  &$6.0^{+0.6}_{-0.5}$&5.85 &6.08 &14.4&1.79\\\hline
15 &93442-01-15-05& 54713.359592  &3681 &3&10.9&$2.62^{+0.13}_{-0.13}$  &$6.6^{+0.6}_{-0.5}$&5.67 &7.92 &14.0&2.00\\\hline
16 &93442-01-16-06 &54717.394303  &4072 &3&9.5 &$3.25^{+0.19}_{-0.17}$  &$4.8^{+0.4}_{-0.4}$&7.01 &7.92 &11.3&1.70\\\hline
17 &93442-01-17-01& 54723.019835  &3875 &3&8.5 &$2.98^{+0.18}_{-0.16}$  &$5.4^{+0.5}_{-0.5}$&6.29 &6.56 &10.4&1.44\\\hline
$18^{*}$ &93442-01-17-05 &54726.704638 & 4430&3&8.0 &$2.41^{+0.10}_{-0.10}$&$8.6^{+0.9}_{-0.8}$&6.92 &8.16 & 11.8 &1.36\\\hline
\end{tabular}
\end{center}
\begin{list}{}{}
\item[${\mathrm{*}}$]{Burst shows photosphere radius expansion.}
\item[Note:]{1 crab is roughly 2500 counts/PCU/s in the 2-60 keV band.
The bolometric flux and fluence are presented in the energy band
1.5-30 keV, and in unit of $erg~cm^{2}~s^{-1}$.
}
\end{list}
\end{table}

\begin{figure}[ptbptbptb]
\centering
 \includegraphics[angle=0, scale=0.5]{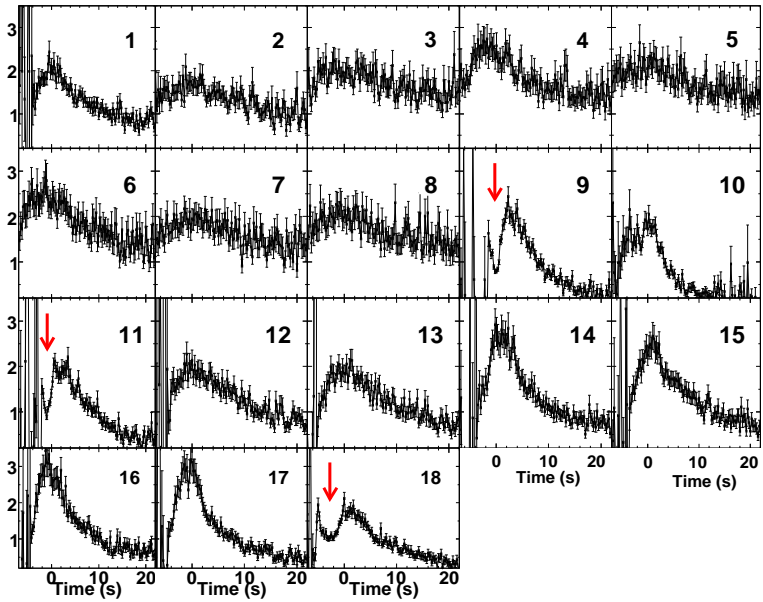}
      \caption{Hardness ratios (6-30 keV)/(2-6 keV) of the 18 bursts. The arrows indicate the PRE bursts.}
         \label{lc_hardness}
\end{figure}

\begin{figure}[ptbptbptb]
\centering
 \includegraphics[angle=0, scale=0.20]{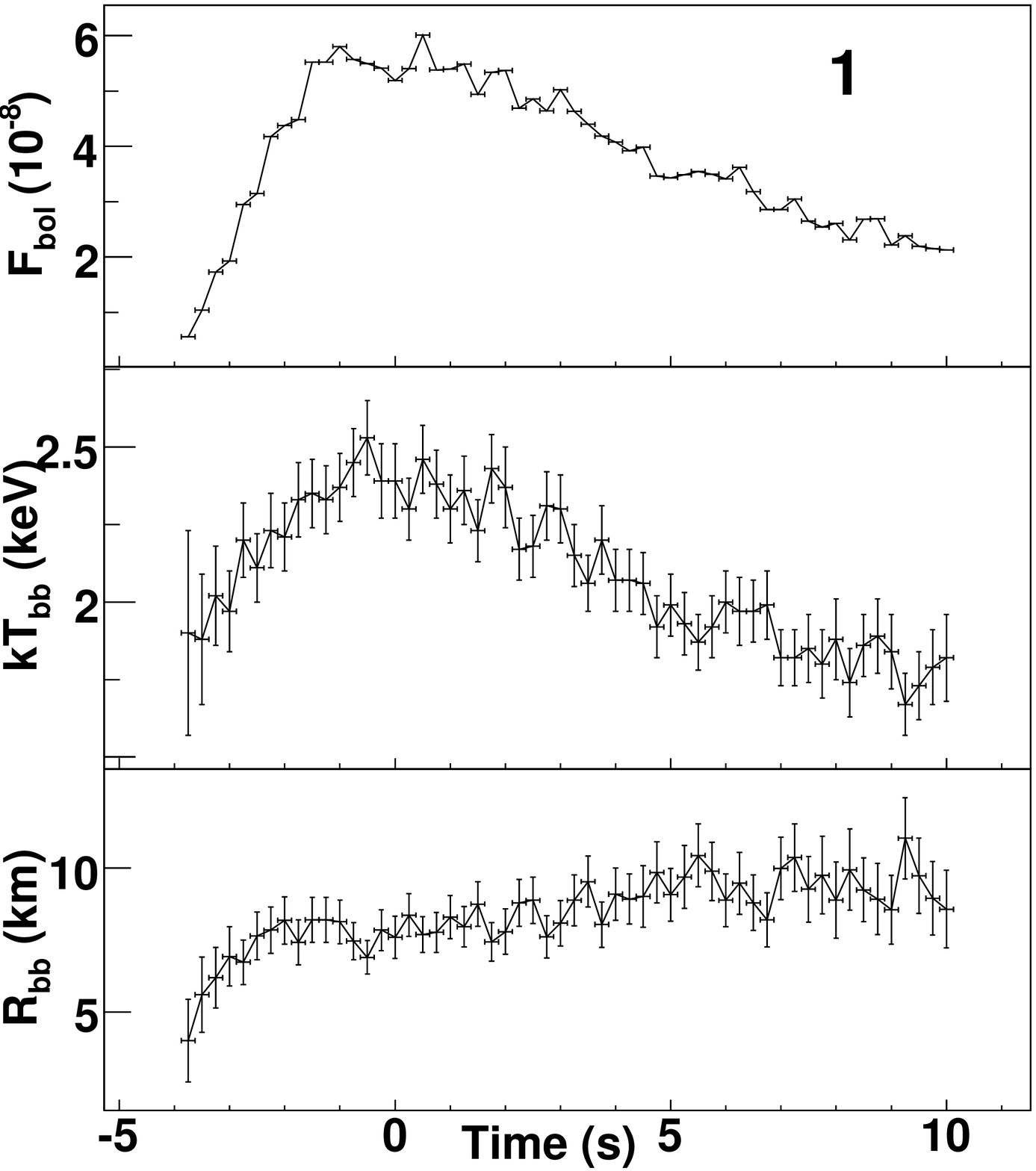}
 \includegraphics[angle=0, scale=0.20]{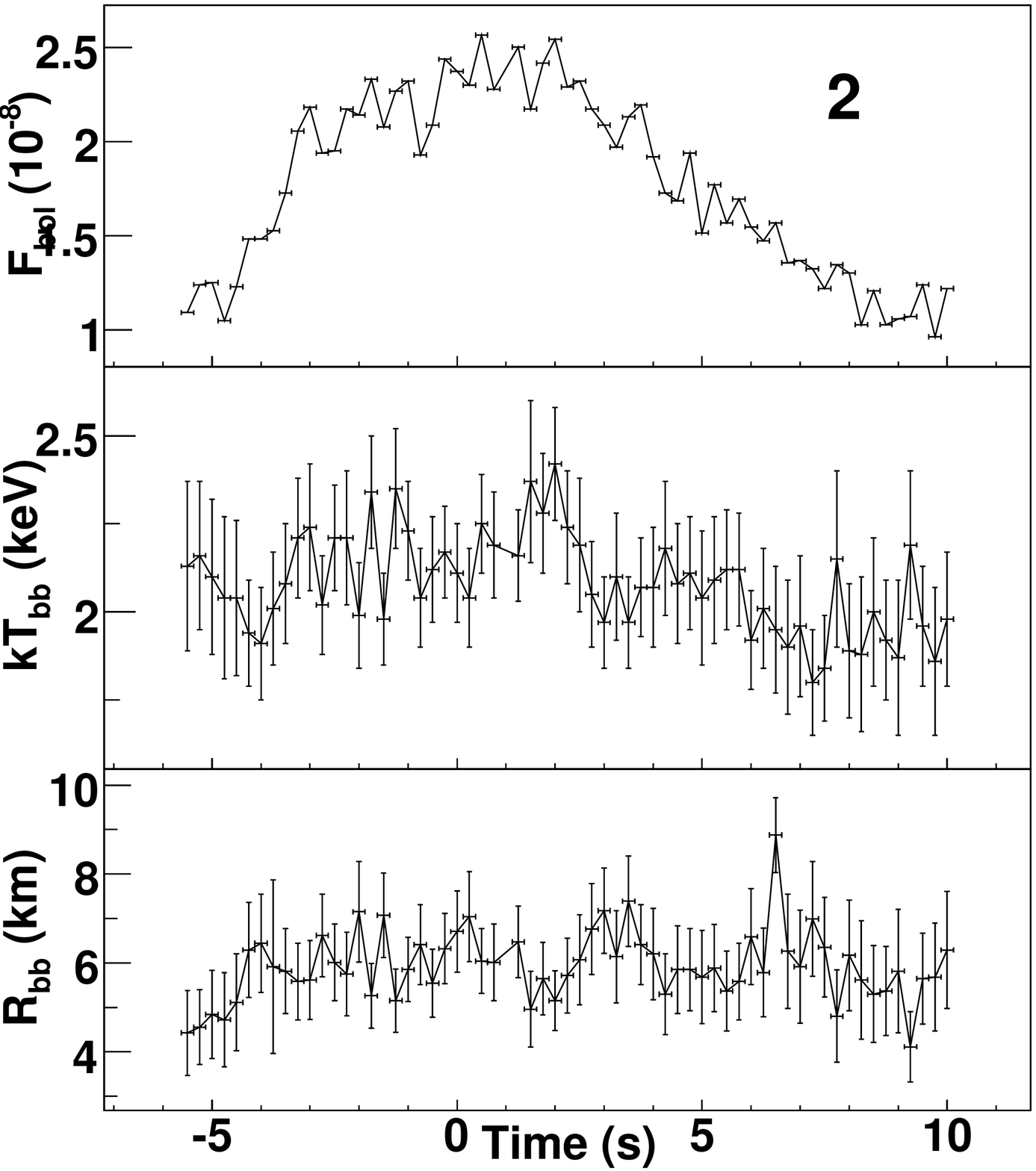}
 \includegraphics[angle=0, scale=0.20]{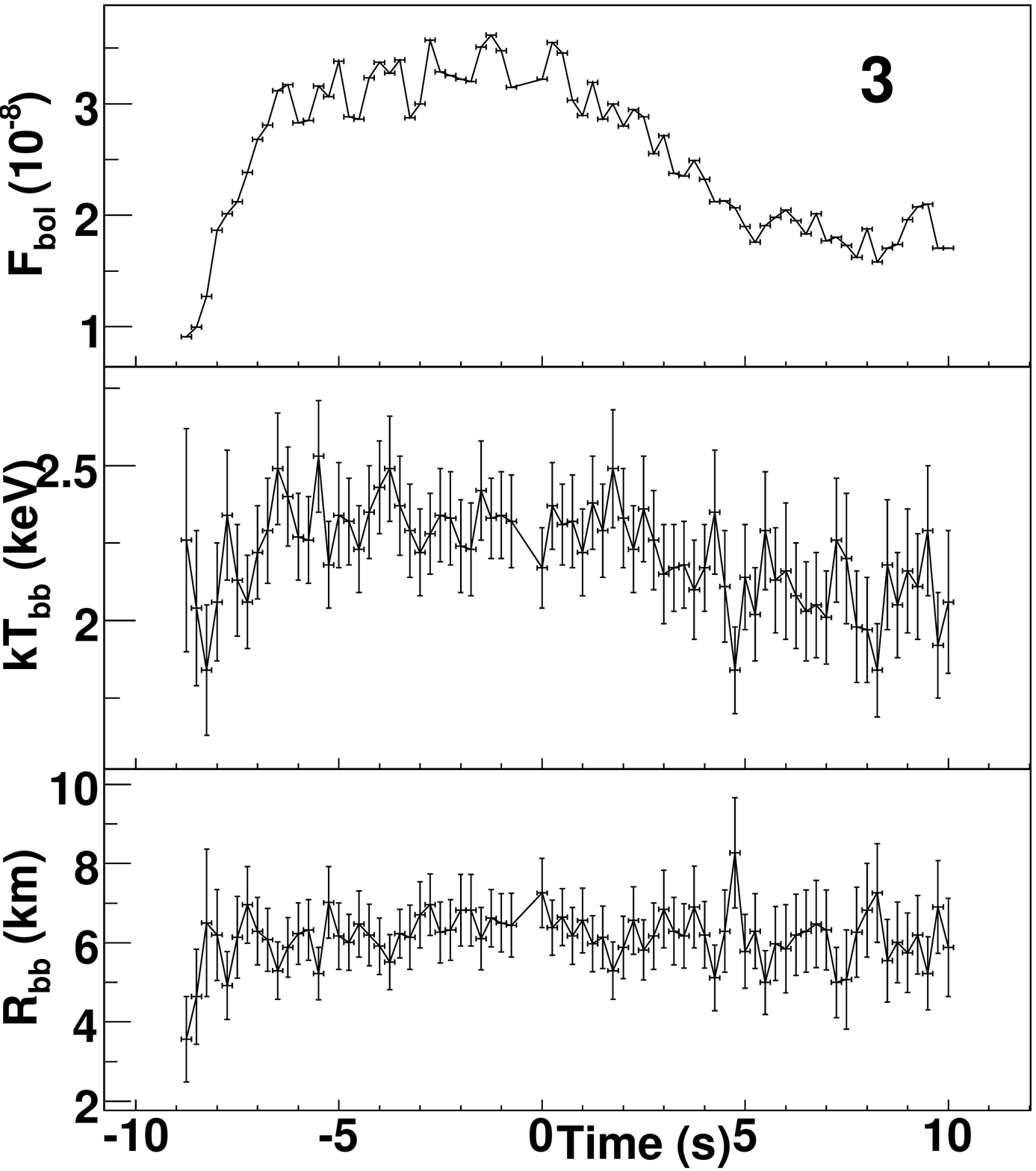}
 \includegraphics[angle=0, scale=0.20]{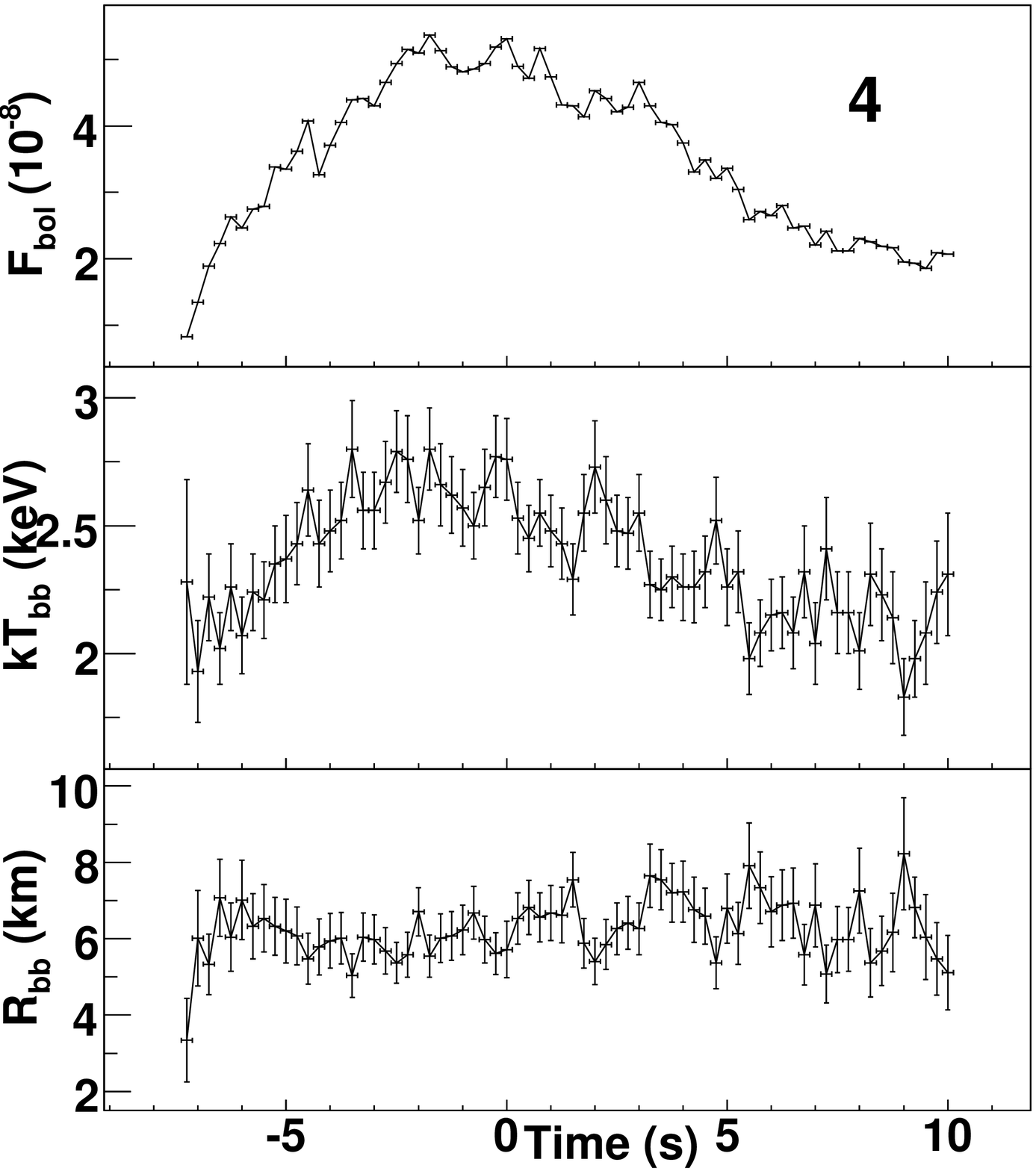}
 \includegraphics[angle=0, scale=0.20]{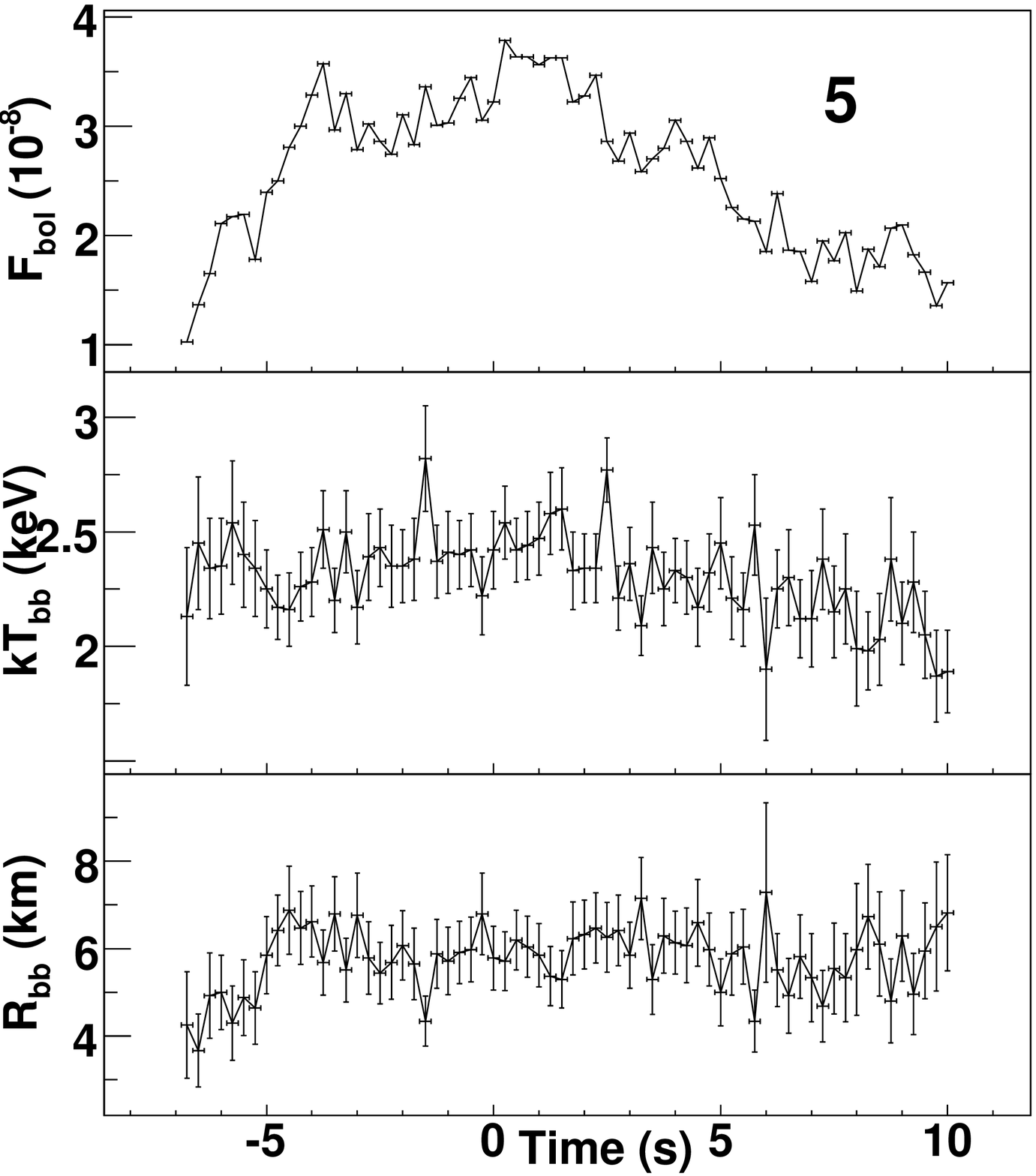}
 \includegraphics[angle=0, scale=0.20]{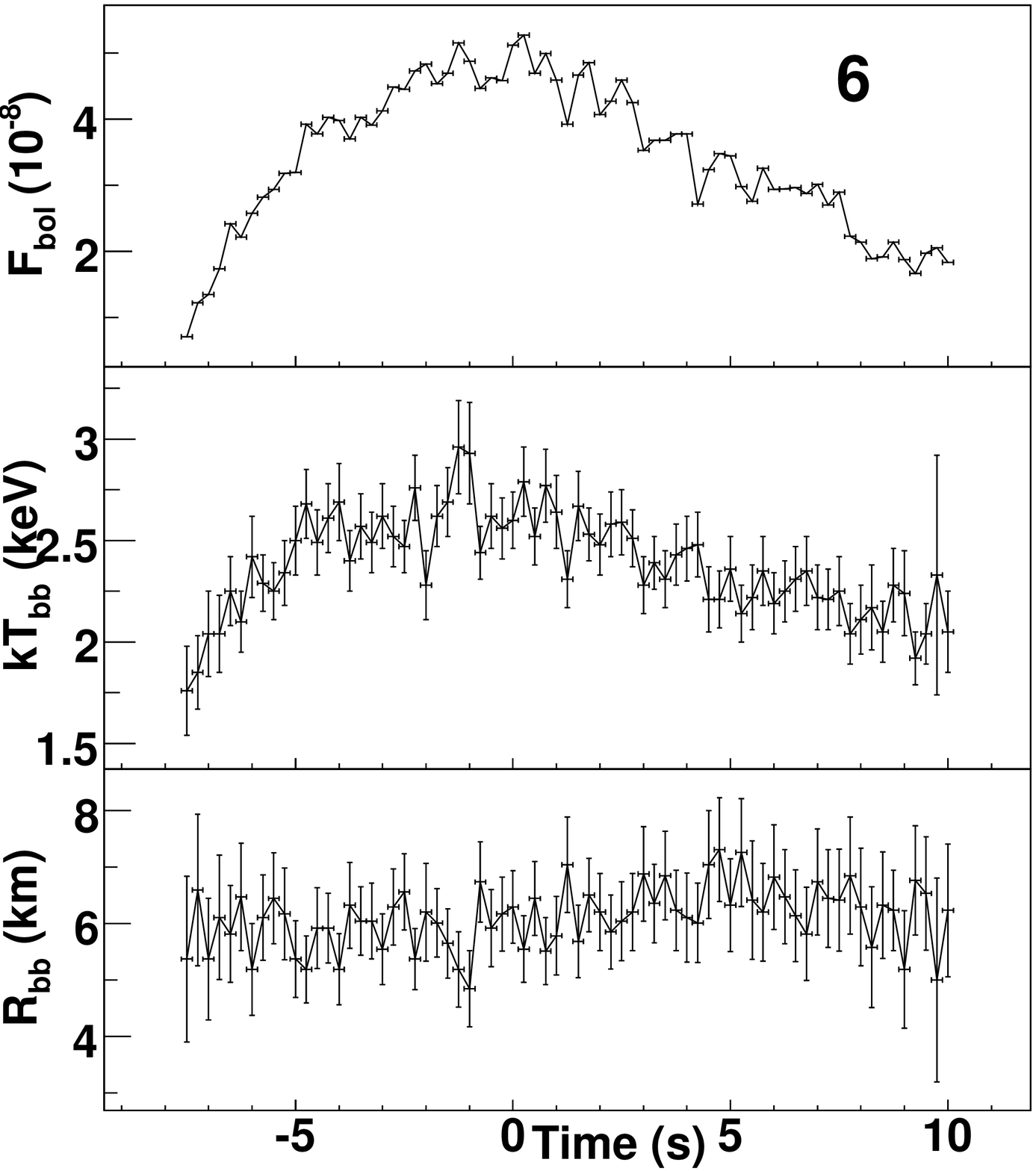}
 \includegraphics[angle=0, scale=0.20]{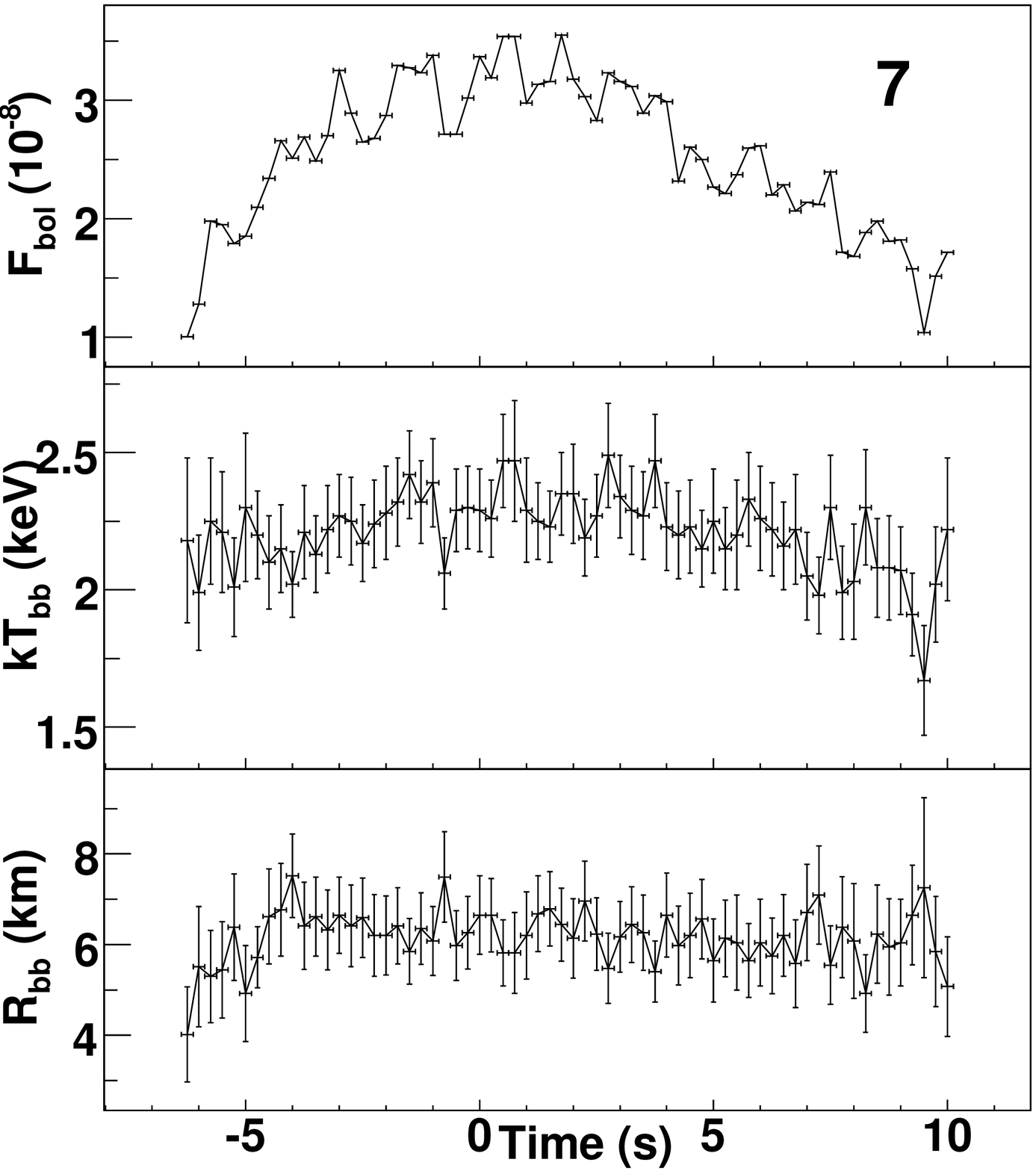}
 \includegraphics[angle=0, scale=0.20]{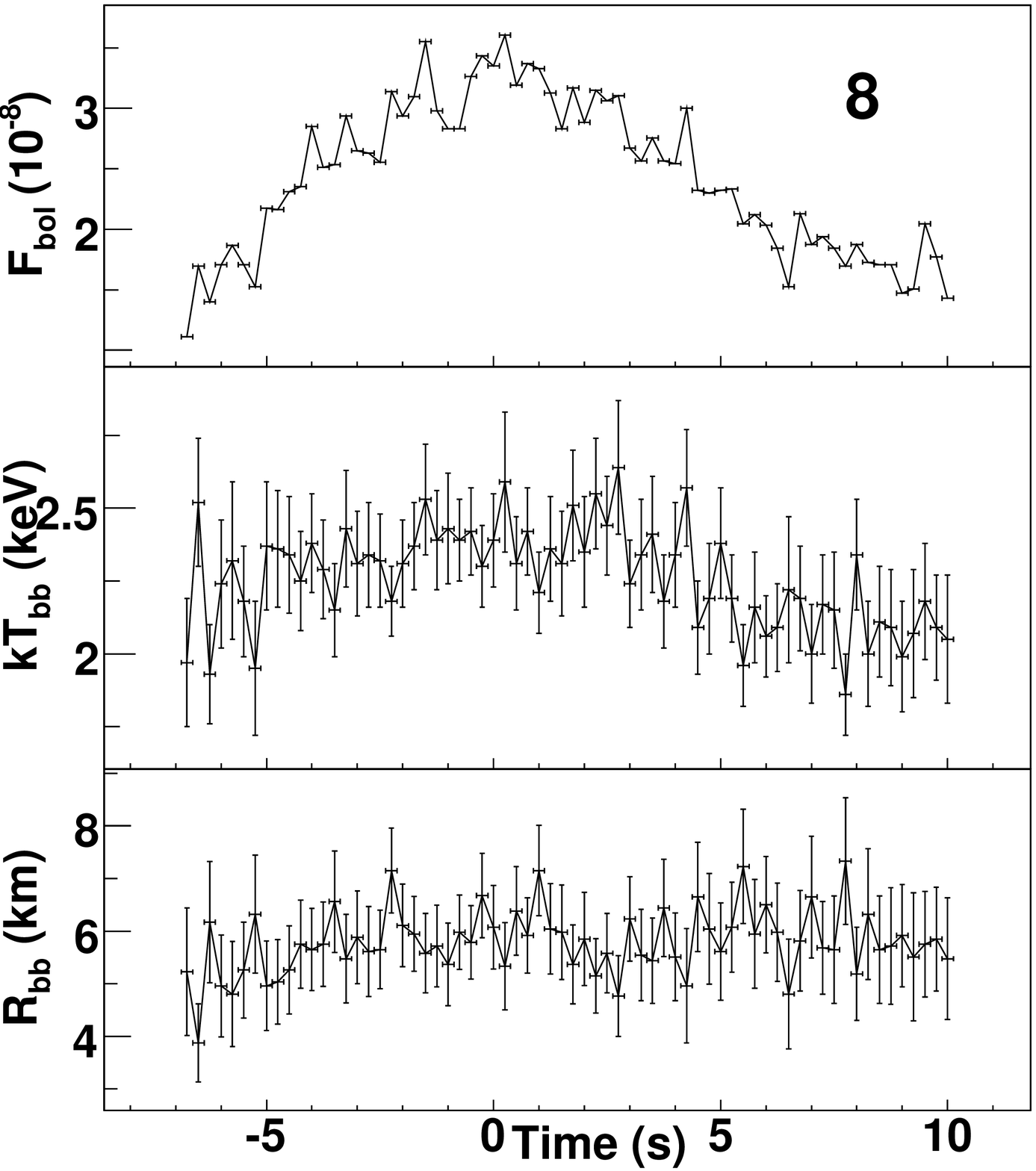}
 \includegraphics[angle=0, scale=0.20]{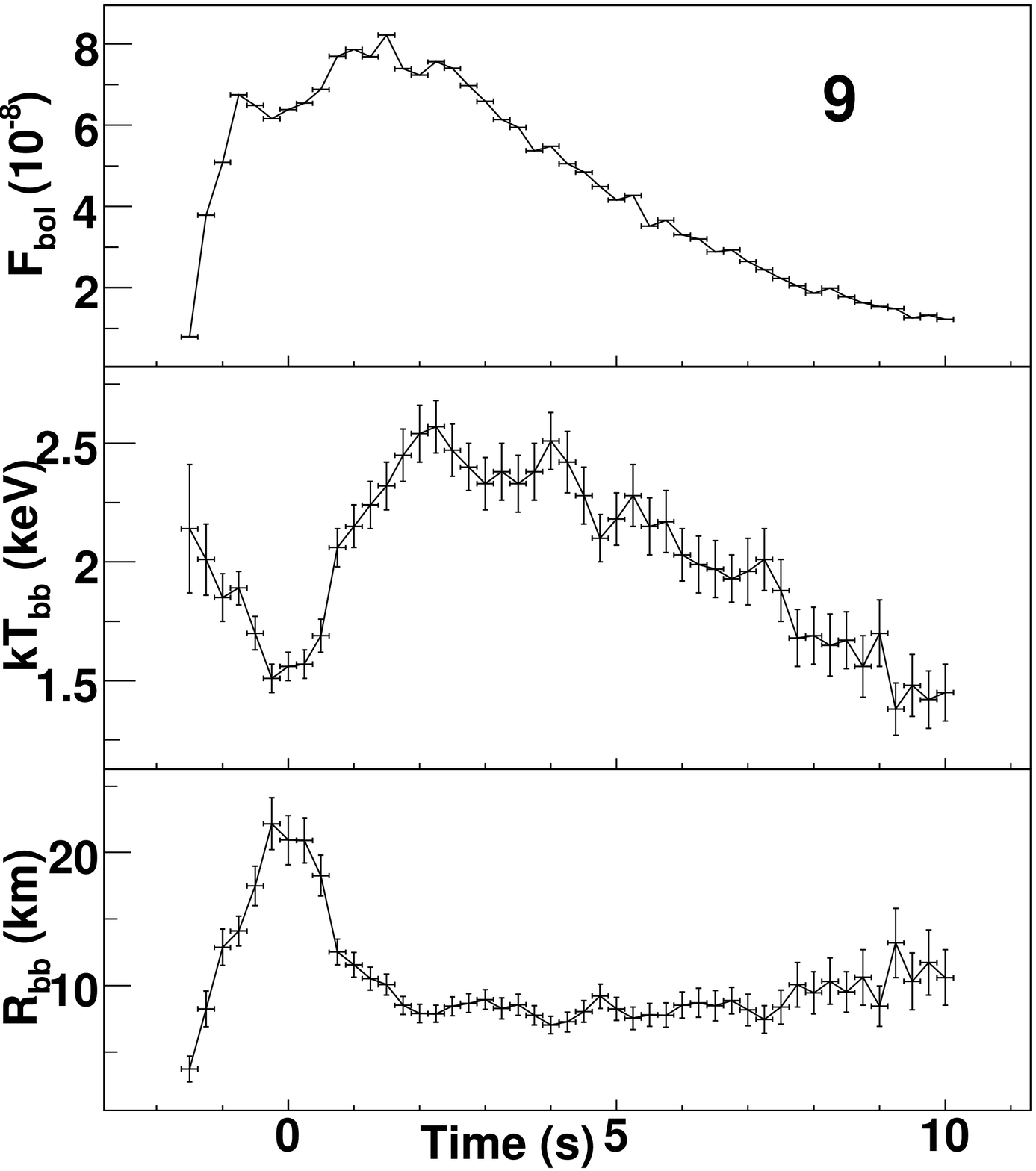}
 \includegraphics[angle=0, scale=0.20]{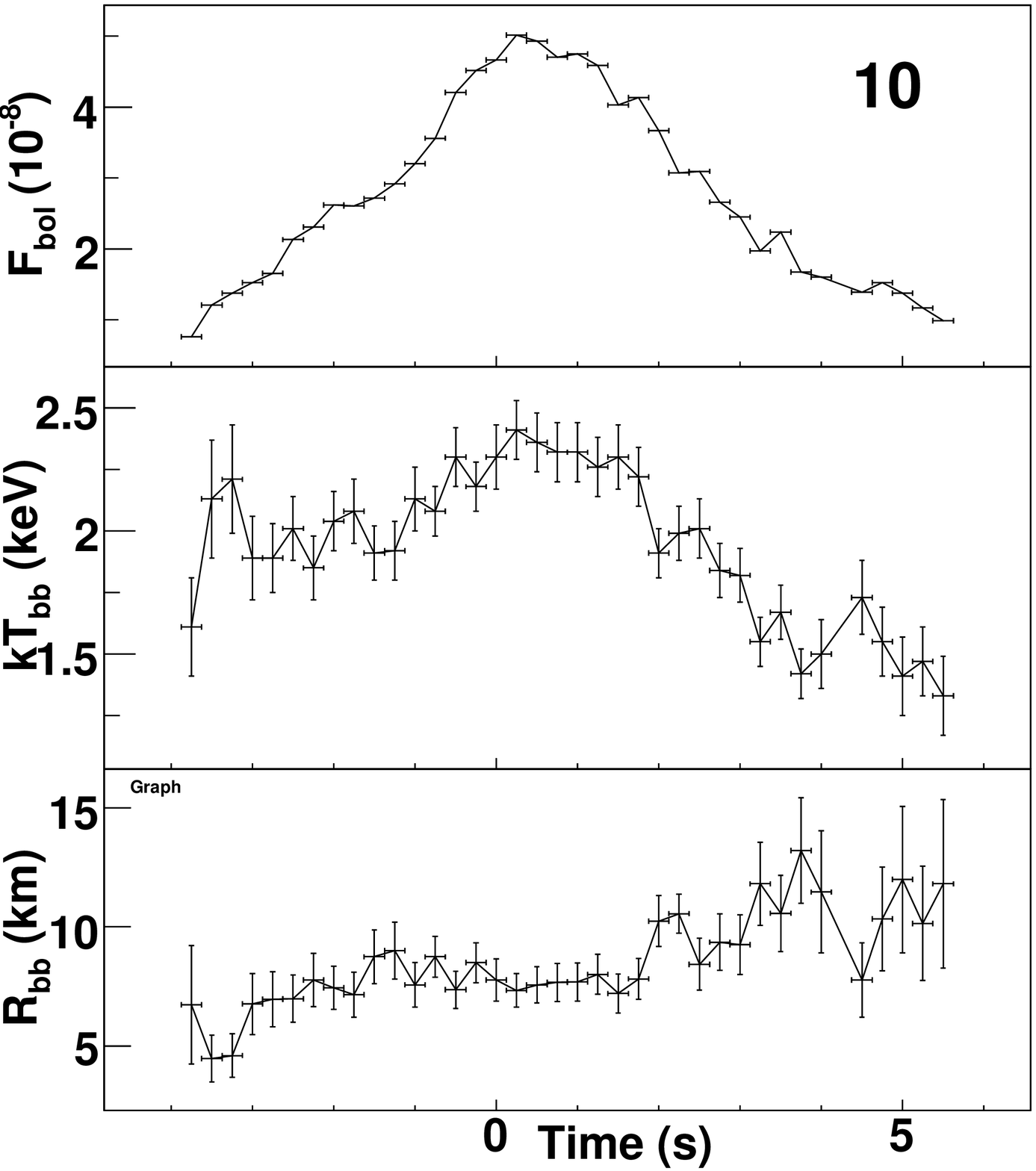}
 \includegraphics[angle=0, scale=0.20]{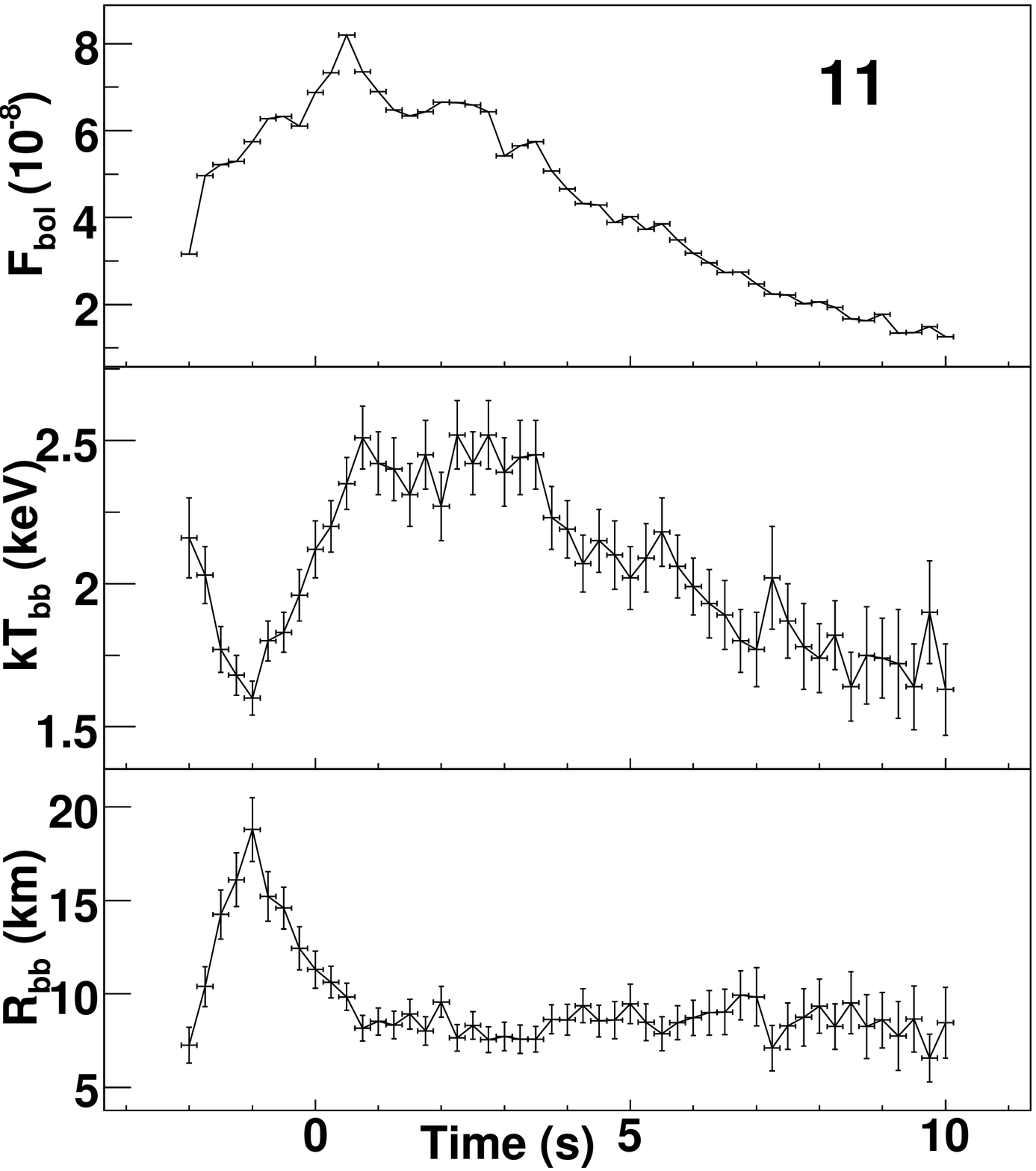}
 \includegraphics[angle=0, scale=0.20]{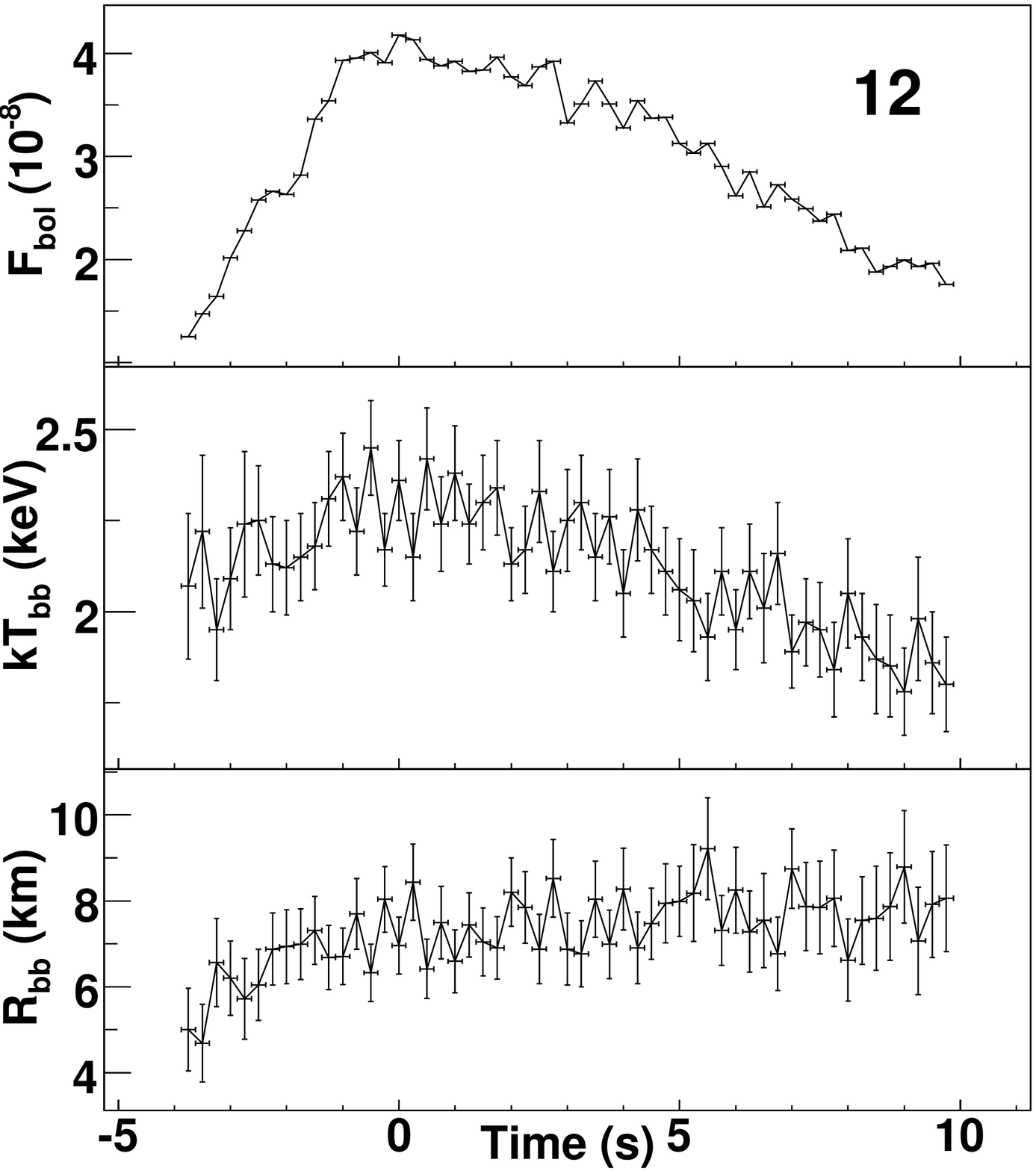}
 \includegraphics[angle=0, scale=0.20]{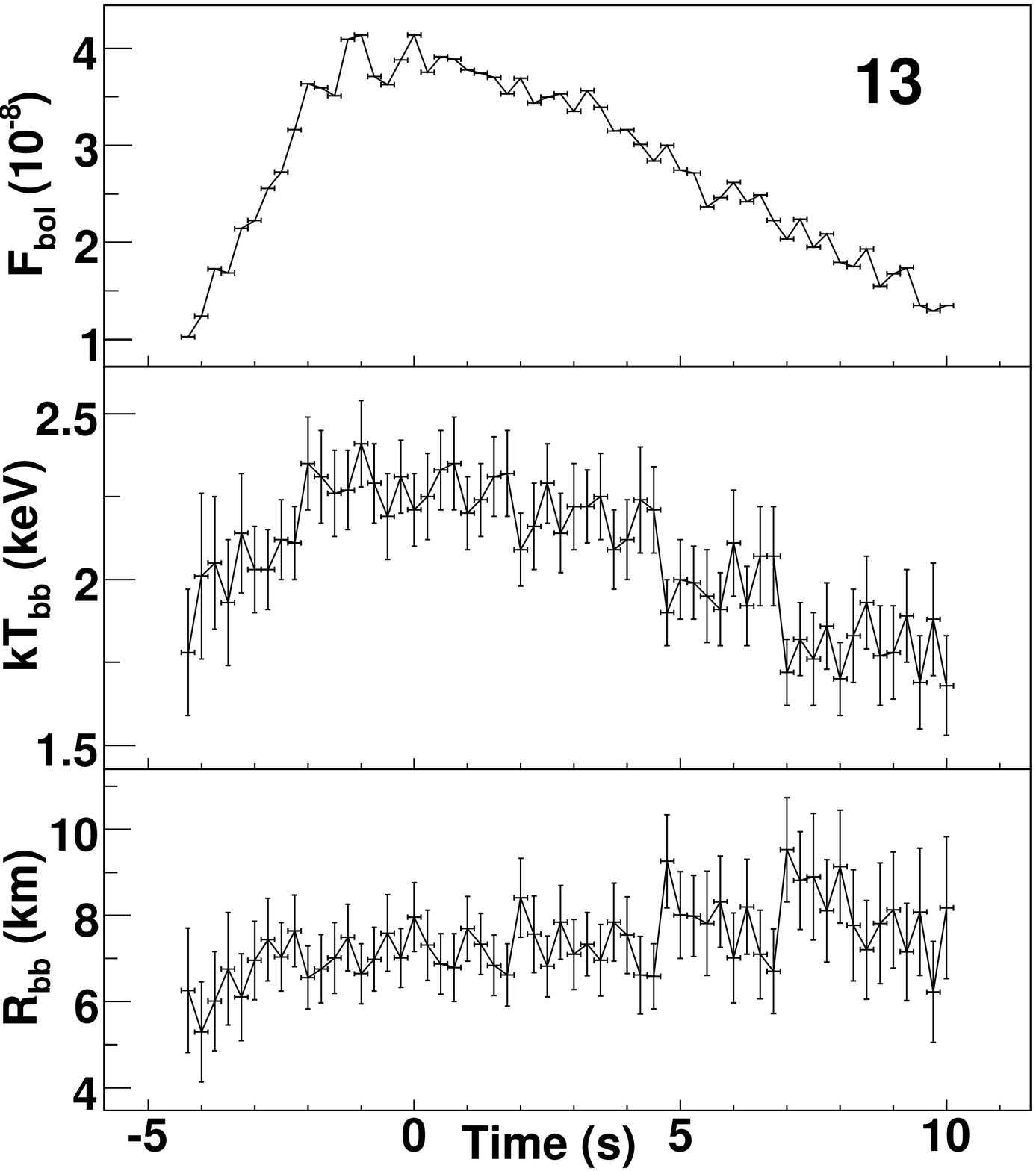}
 \includegraphics[angle=0, scale=0.20]{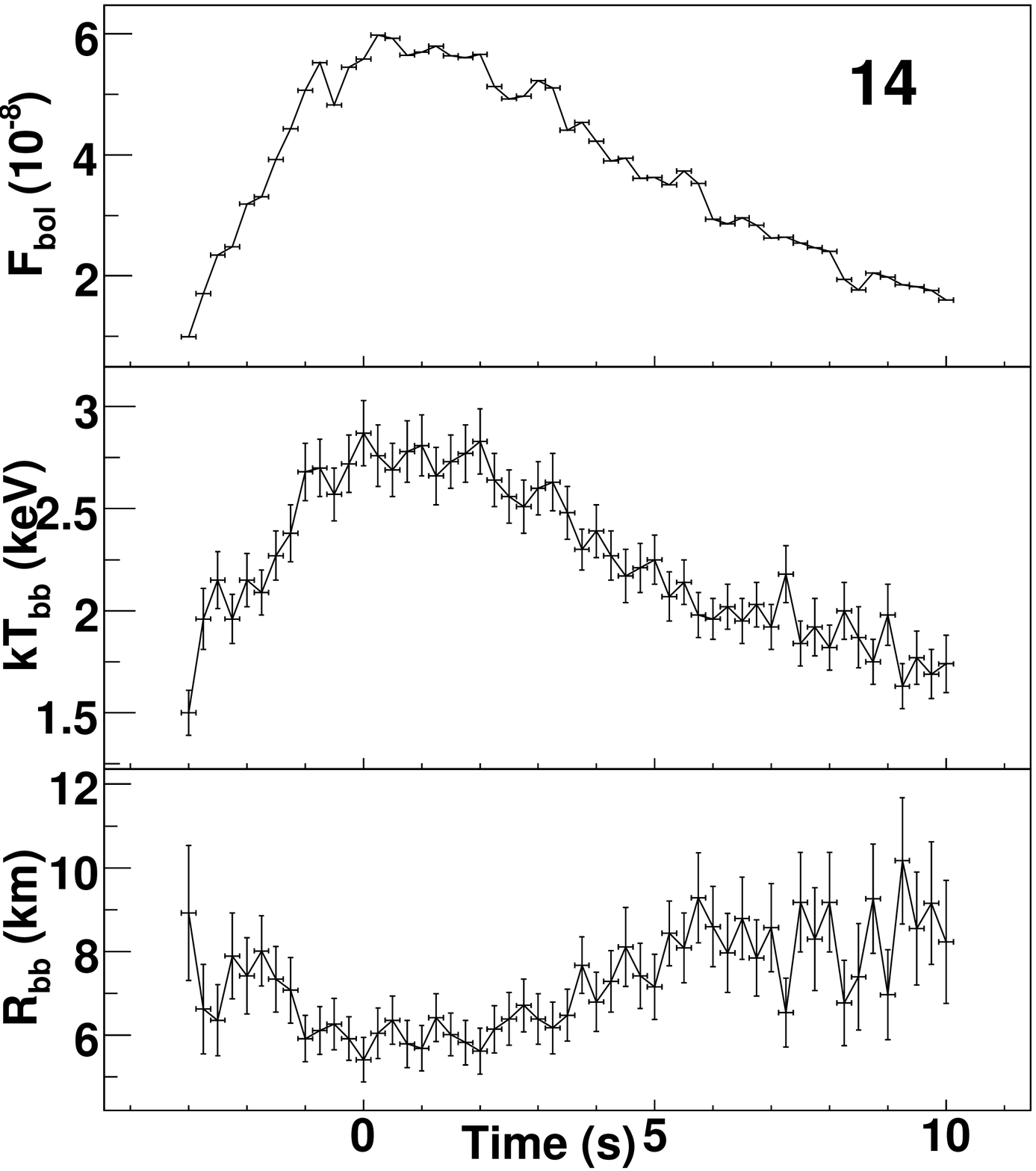}
 \includegraphics[angle=0, scale=0.20]{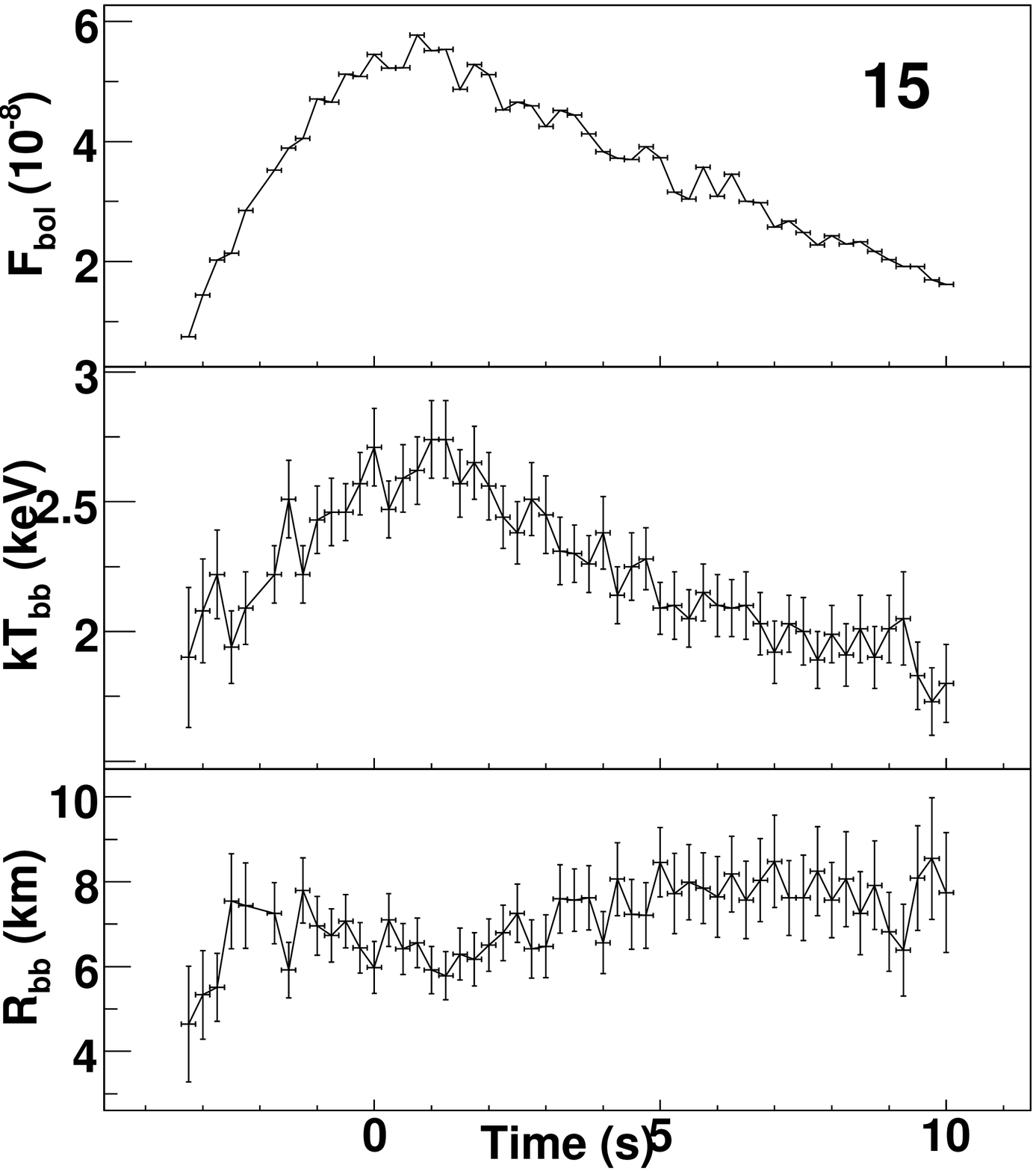}
 \includegraphics[angle=0, scale=0.20]{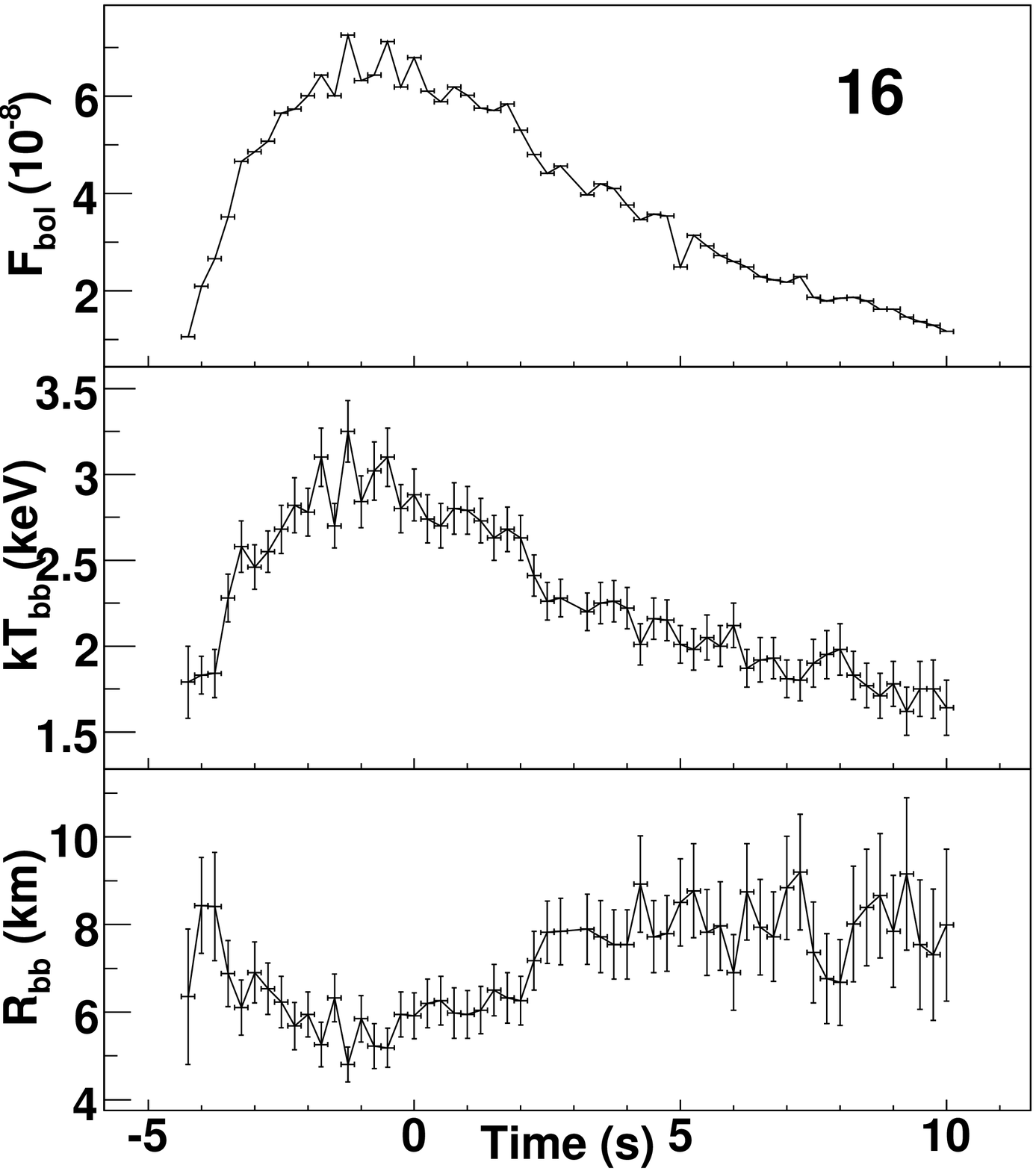}
 \includegraphics[angle=0, scale=0.20]{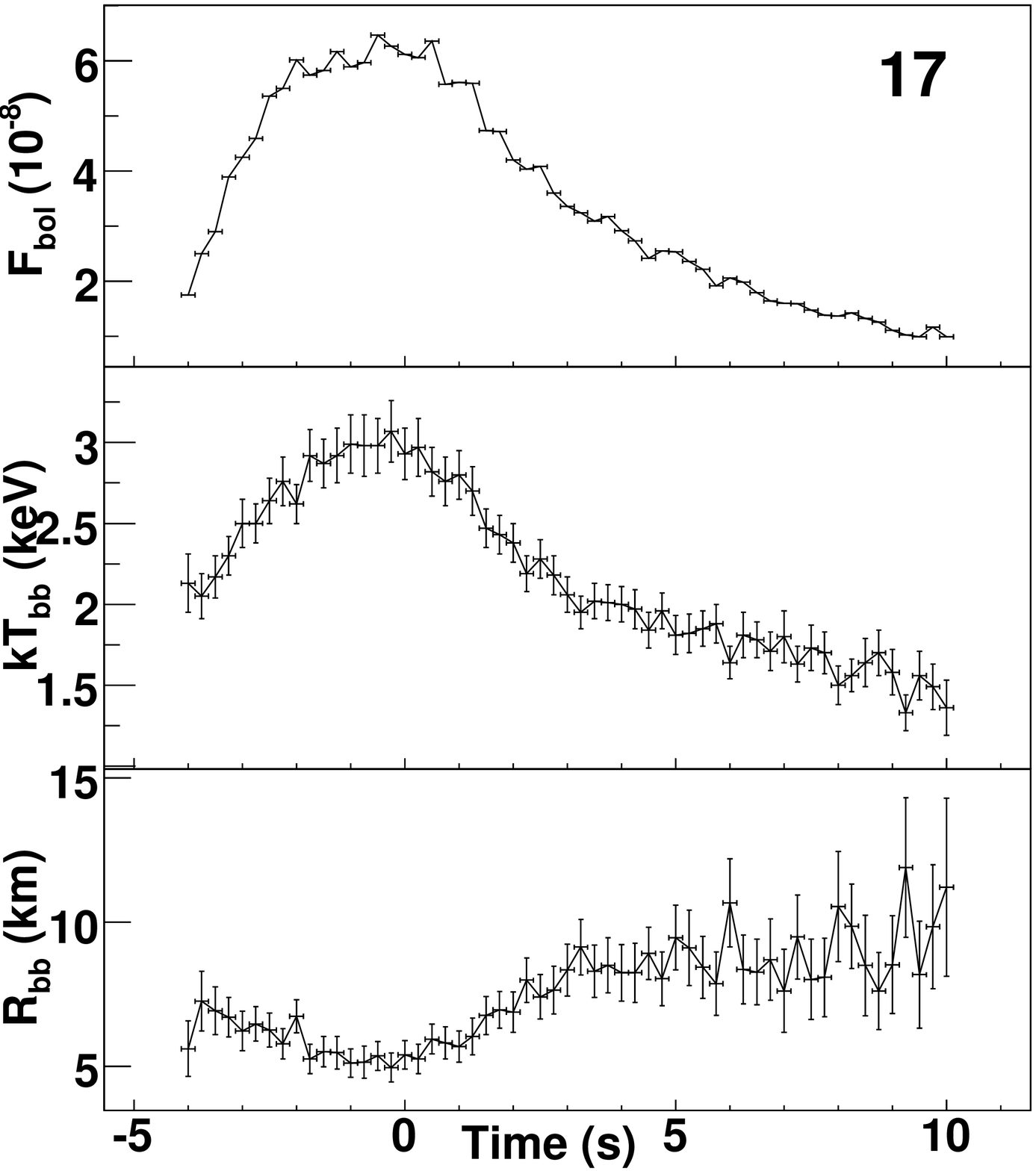}
 \includegraphics[angle=0, scale=0.20]{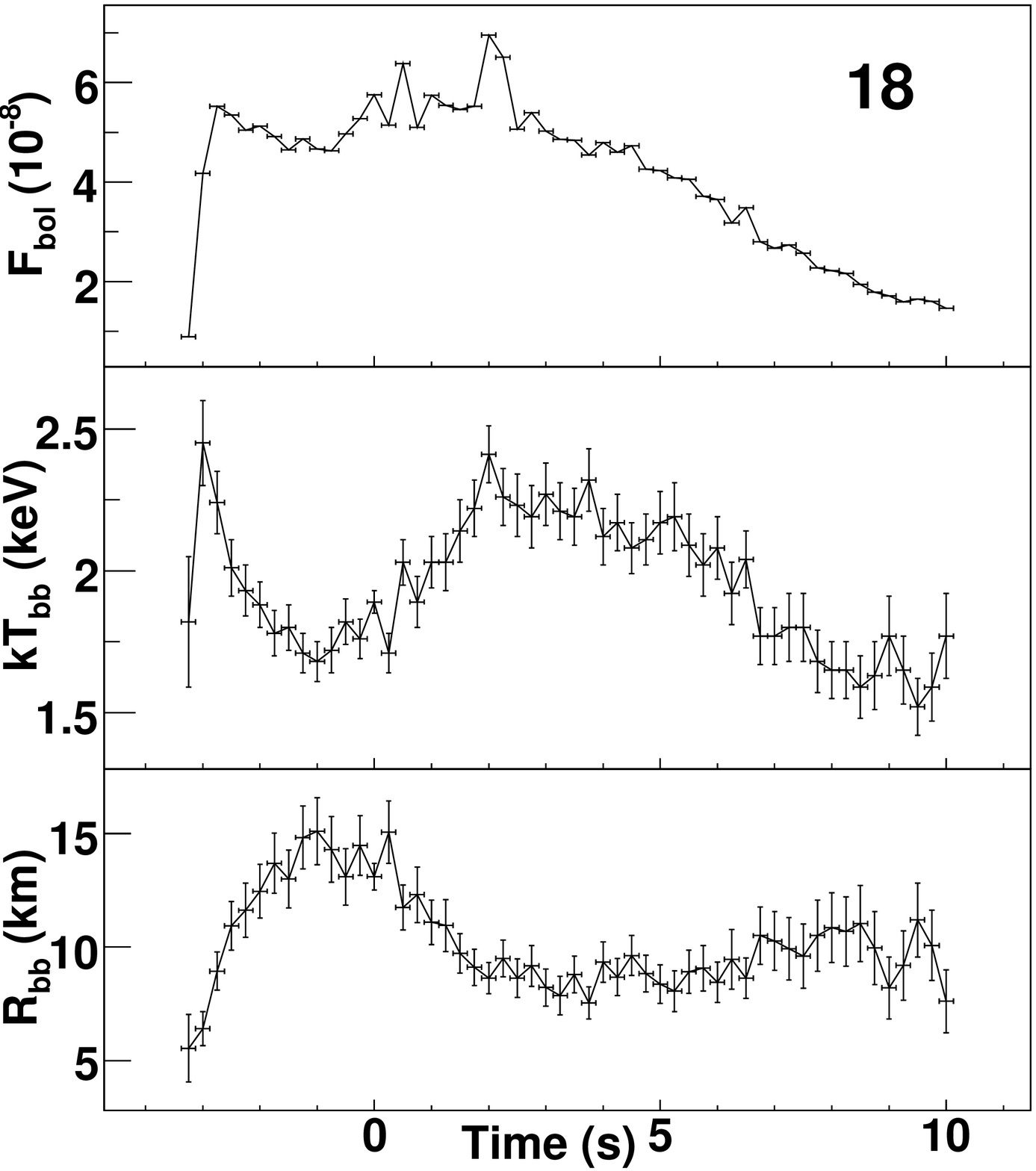}
      \caption{The evolution of the burst luminosity, temperature, and radius from fitting the data with a blackbody model, for the 18 bursts detected in IGR J17473-2721.  The time zero in the X-axis is defined as when intensity reaches the peak  (see Table 1). F$_{bol}$ is the bolometric flux in units of erg cm$^{-2}$s$^{-1}$.}
         \label{lc_lum}
\end{figure}

\begin{figure}[ptbptbptb]
\centering
 \includegraphics[angle=0, scale=0.32]{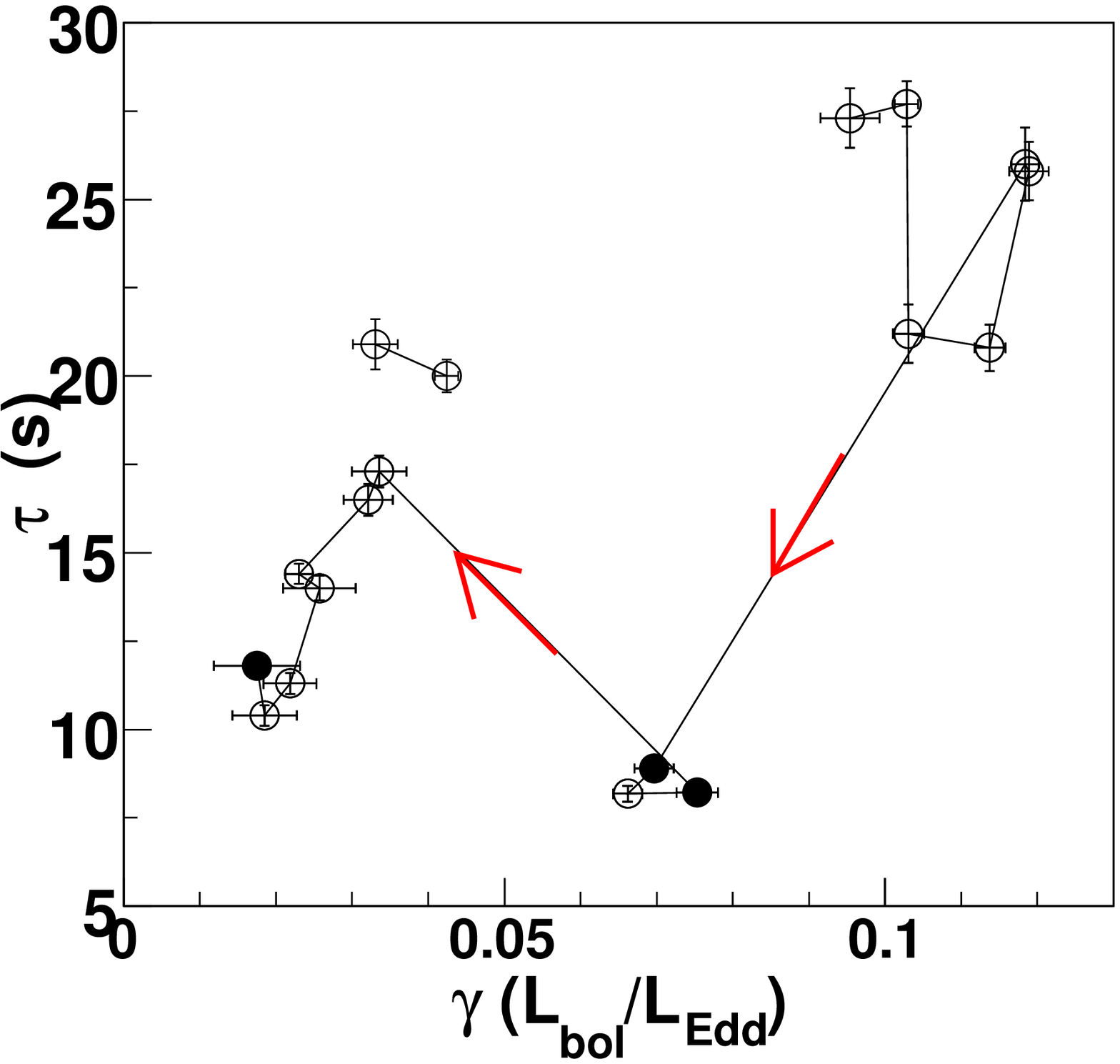}
  \includegraphics[angle=0, scale=0.32]{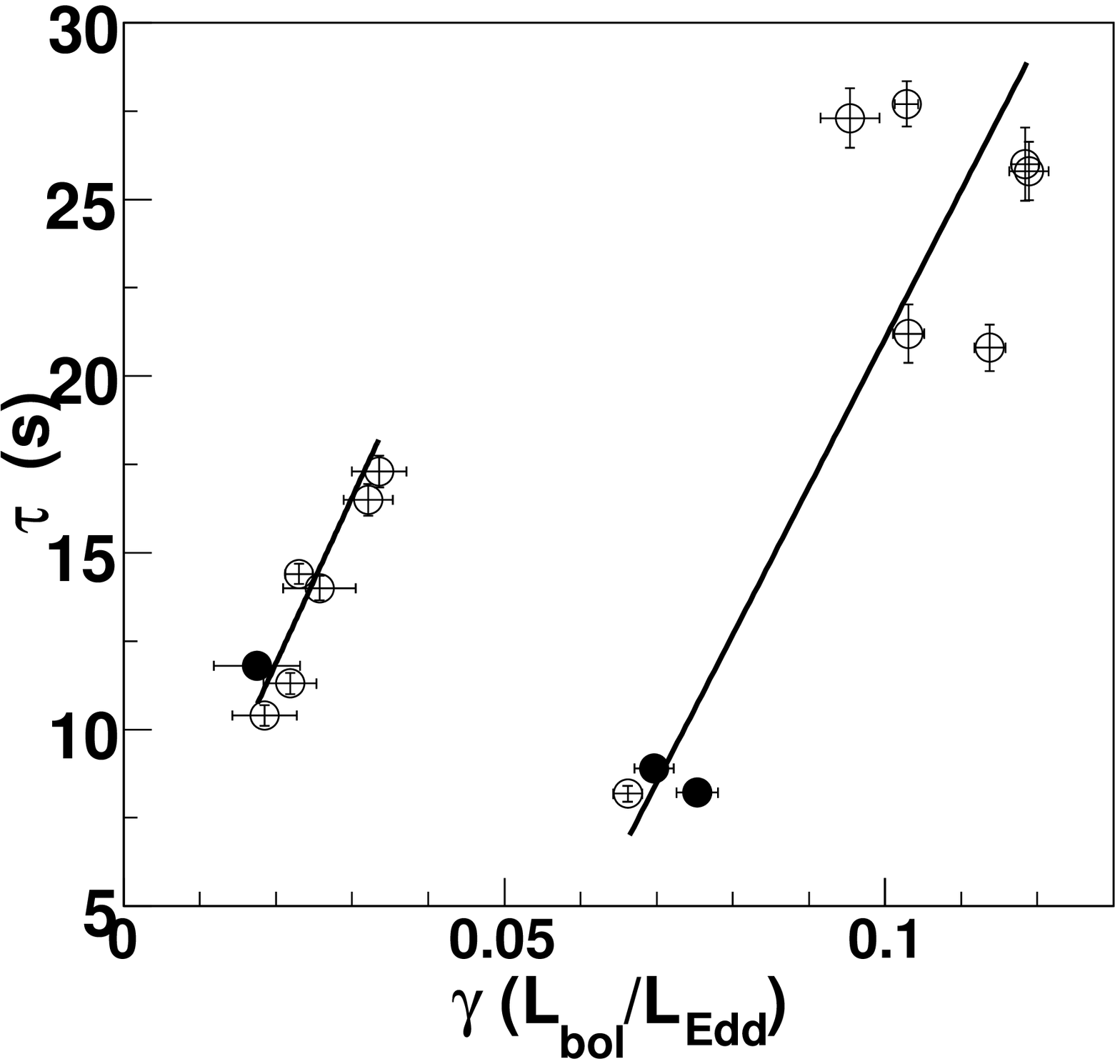}
      \caption{Characteristic  timescale $\tau$ vs the Eddington ratio $\gamma$. The left panel shows the sequence along the two outbursts, and the right panel shows a linear fit to the two groups of bursts. The filled circles are the three PRE bursts and the open circles the others.}
         \label{f_per-tal}
\end{figure}

\begin{figure}[ptbptbptb]
\centering
  \includegraphics[angle=0, scale=0.5]{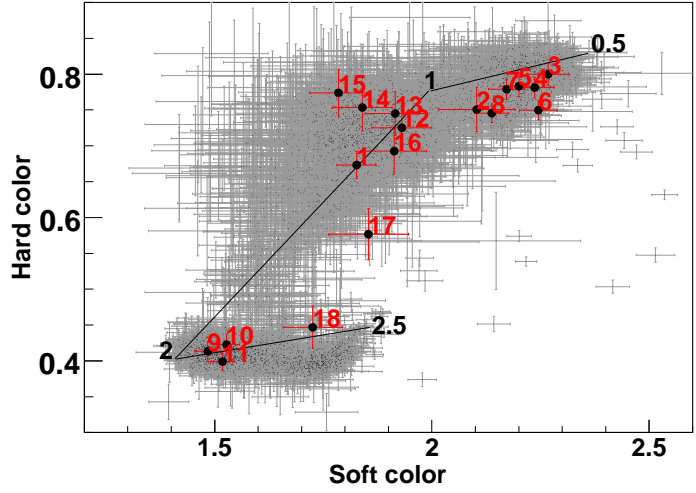}

      \caption{The color-color diagram of IGR~J17473-2721. The integration time is 64 seconds per point.  The  points close to each burst are labeled with a filled circle and a red number. The black solid curve is the atoll-track  used to calculate $S_{a}$ (see text), and the locations of $S_{a}$=0.5, 1, 2 and 2.5 are labeled with black characters. }
         \label{burst_color_color}
\end{figure}

\begin{figure}[ptbptbptb]
\centering
  \includegraphics[angle=0, scale=0.5]{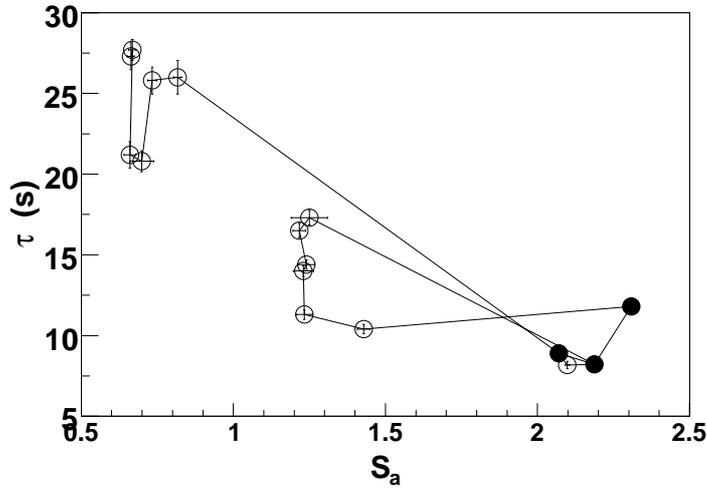}
      \caption{Characteristic timescale $\tau$ vs the $S_{a}$. The filled circles are the three PRE bursts and the open circles the others.}
         \label{burst_Sz}
\end{figure}

\begin{figure}[ptbptbptb]
\centering
 \includegraphics[angle=0, scale=0.5]{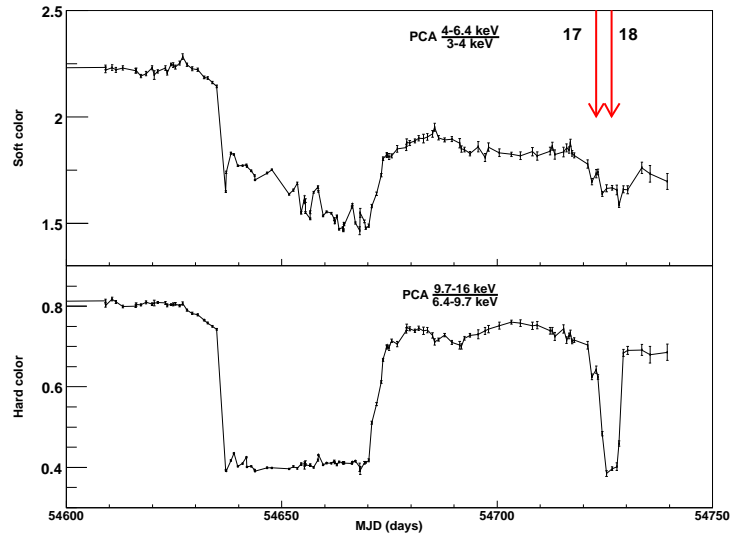}
      \caption{The evolution of the hardness of the second outburst. The upper panel shows the flux ratio of (4-6.4 keV)/(3-4 keV), whereas the lower panel  shows the flux ratio of (9.7-16 keV)/(6.4-9.7 keV). The data points are averaged over each  OBSID of typical $\sim$ 3000 second durations. The locations of burst \# 17 and \# 18 are marked by arrows.
        }
         \label{burst_hardness}
\end{figure}

\begin{figure}[ptbptbptb]
\centering
  \includegraphics[angle=0, scale=0.5]{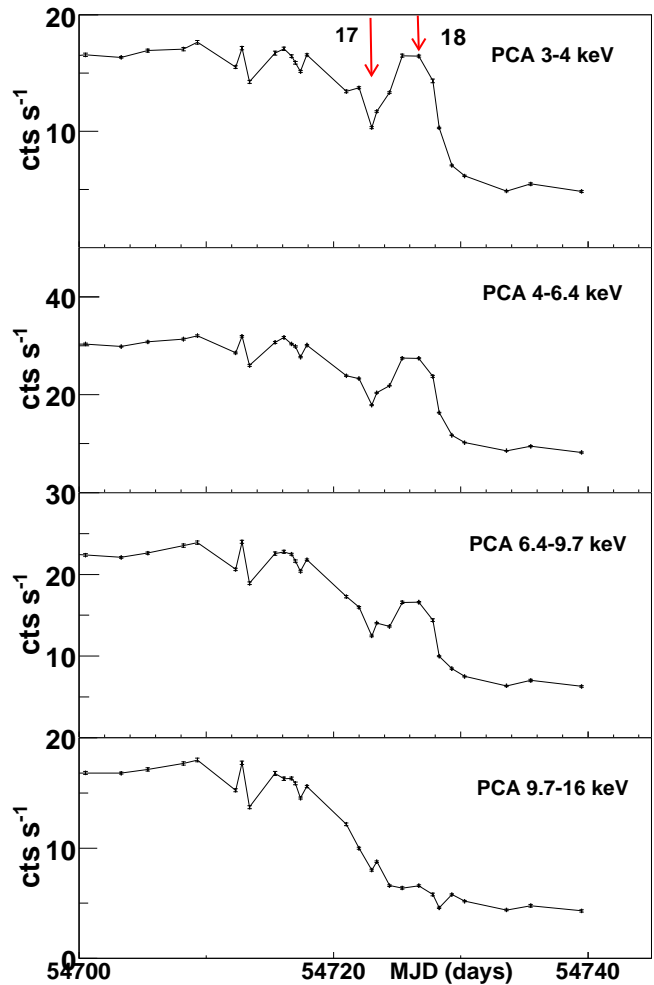}
      \caption{The flux evolutions of the hard X-ray dip during the decay of the second outburst in four  energy bands.  The data points are averaged over each  OBSID of typical $\sim$ 3000 second duration.  The location of burst \# 17 and \# 18 are marked by arrows.
        }
         \label{burst_hardness_1}
\end{figure}


\begin{thebibliography}{}



\bibitem[Altamirano et al. (2008a)]{altamirano08a}Altamirano D., Degenaar N., and Markwardt C. 2008a, ATel, 1459

\bibitem[Altamirano et al. (2008b)]{altamirano08b}Altamirano D., Galloway D., Chenevez J., et al. 2008b, ATel, 1651

\bibitem[Basinska et al. (1984)]{Basinska}Basinska E. M., Lewin W. H. G., Sztajno M., et al. 1984, ApJ, 281, 337

\bibitem[Belian et al. (1976)]{Belian}Belian R. D., Conner J. P., \& Evans W. D. 1976, ApJ, 206, L135

\bibitem[Bhattacharyya et al. (2009)]{Bha}Bhattacharyya1 S., Miller M. C., \& Galloway D. K. 2009, MNRAS, inpress (arXiv:0908.4245)

\bibitem[Bildsten (2000)]{Bli2000}Bildsten L. 2000, AIPC, 522, 359

\bibitem[Cumming et al. (2004)]{Cumming}Cumming A. 2004, Nuclear Physics B Proceedings Supplements, 132, 435

\bibitem[Del Monte et al. (2008)]{monte08}Del Monte E., Evangelista Y., Feroci M., et al. 2008, ATel, 1445

\bibitem[Fujimoto et al. (1981)]{Fuji}Fujimoto M. Y., Hanawa T., \& Miyaji S. 1981, ApJ, 247, 267

\bibitem[Galloway et al. (2008)]{Galloway}Galloway D. K., Muno M. P., Hartman, et al. 2008, ApJS, 179, 360

\bibitem[Grindlay et al. (1976)]{Grindlay}Grindlay J., Gursky H., Schnopper H., et al. 1976, ApJ, 205, L127

\bibitem[Hasinger et al. (1989) ]{hasinger89}Hasinger G., \& van der Klis M. 1989, A\&A, 225, 79

\bibitem[in 't Zand et al. (2009)]{int}in't Zand J. J. M., Keek L., Cumming A. et al. 2009, A\&A, 497, 469

\bibitem[Kuulkers et al. (2003)]{Kuulkers}Kuulkers E., den Hartog, P. R., in¡¯t Zand, J. J. M., et al. 2003, A\&A, 399, 663


\bibitem[Lewin et al. (1993)]{Lewin}Lewin W. H. G., van Paradijs J., \& Taam R. E. 1993, Space Science Reviews, 62, 223

\bibitem[M\'{e}ndez et al. (1999)]{Mend}M\'{e}ndez M., van der Klis M., Ford E. C. et al. 1999, ApJ, 511, L49

\bibitem[Muno et al. (2000)]{Muto}Muno M. P., Fox D. W., Morgan E. H. et al. 2000, ApJ, 542, 1016

\bibitem[Strohmayer et al. (2006)]{Strohmayer}Strohmayer T. \& Bildsten L. 2006, New views of thermonuclear bursts (Compact stellar X-ray sources), 113¨C156

\bibitem[van der Klis (1995)]{van1995}van der Klis, M. 1995, in X-ray Binaries, p. 252, Lewin, W. H. G., Van Paradijs, J., \& Van den Heuvel, E. P. J. (Eds), Cambridge University Press

\bibitem[van der Klis et al. (2006)]{van der Klis}
            van der Klis, M. 2006, in Compact Stellar X-ray Sources (eds. W. Lewin \& M. van der Klis, Cambridge University Press), 39¨C112

\bibitem[Weinberg et al. (2006)]{Weig}Weinberg N. N., Bildsten L., \& Schatz h., 2006, ApJ, 639, 1018

\bibitem[Zhang et al. (2009)]{zhanggb09}Zhang G.B., Mendez M., Altamirano D., et al., 2009, MNRAS, 398, 368

\bibitem[Zhang G. B. et al. (2009)]{zhangshu09}Zhang S., Chen Y.P., Wang J.M., et al. 2009, A\&A, 502, 231





\end{thebibliography}
\end{document}